\documentclass[fleqn,usenatbib]{mnras}

\usepackage{newtxtext,newtxmath,gensymb,bm}

\usepackage[T1]{fontenc}

\DeclareRobustCommand{\VAN}[3]{#2}
\let\VANthebibliography\thebibliography
\def\thebibliography{\DeclareRobustCommand{\VAN}[3]{##3}\VANthebibliography}

\usepackage{graphicx}	
\usepackage{amsmath}	
\usepackage{pdflscape}
\usepackage{afterpage}

\newcommand{\Eq}[1]{Eq.~\ref{#1}}
\newcommand{\Fig}[1]{Fig.~\ref{#1}}
\newcommand{\Sec}[1]{Section~\ref{#1}}
\newcommand{\App}[1]{Appendix~\ref{#1}}
\newcommand{\Tab}[1]{Table~\ref{#1}}
\newcommand{\threehun}{{\sc The Three Hundred}}
\newcommand{\tth}{{\sc The300}}

\title[Dynamical state of clusters]{Reconsidering the dynamical states of galaxy clusters using PCA and UMAP}

\author[R. Haggar et al.]{{Roan Haggar,$^{1,2}$\thanks{E-mail: rhaggar@uwaterloo.ca}
Federico De Luca,$^{3,4}$
Marco De Petris,$^{5}$
Elizaveta Sazonova,$^{1,2}$
James E. Taylor,$^{1,2}$}
\newauthor{Alexander Knebe,$^{6,7,8}$
Meghan E. Gray,$^{9}$
Frazer R. Pearce,$^{9}$
Ana Contreras-Santos,$^{6}$
Weiguang Cui,$^{6,7,10}$}
\newauthor{Ulrike Kuchner,$^{9}$
Robert A. Mostoghiu Paun,$^{11,12}$
Chris Power$^{8,13}$}
\\
$^{1}$Department of Physics and Astronomy, University of Waterloo, Waterloo, Ontario N2L 3G1, Canada\\
$^{2}$Waterloo Centre for Astrophysics, University of Waterloo, Waterloo, Ontario N2L 3G1, Canada\\
$^{3}$Dipartimento di Fisica, Universit\`{a} di Roma Tor Vergata, Via della Ricerca Scientifica 1, I-00133 Roma, Italy\\
$^{4}$INFN, Sezione di Roma 2, Universit\`{a} degli studi di Roma Tor Vergata, Via della Ricerca Scientifica, 1, Roma, Italy\\
$^{5}$Dipartimento di Fisica, Sapienza Universit\`{a} di Roma, Piazzale Aldo Moro 5, I-00185 Roma, Italy\\
$^{6}$Departamento de F\'isica Te\'{o}rica, M\'{o}dulo 15 Universidad Aut\'{o}noma de Madrid, 28049 Madrid, Spain\\
$^{7}$Centro de Investigaci\'{o}n Avanzada en F\'{i}sica Fundamental (CIAFF), Universidad Aut\'{o}noma de Madrid, 28049 Madrid, Spain\\
$^{8}$International Centre for Radio Astronomy Research, University of Western Australia, 35 Stirling Highway, Crawley, Western
Australia 6009, Australia\\
$^{9}$School and Physics \& Astronomy, University of Nottingham, Nottingham NG7 2RD, UK\\
$^{10}$Institute for Astronomy, University of Edinburgh, Royal Observatory, Edinburgh EH9 3HJ, UK\\
$^{11}$Centre for Astrophysics \& Supercomputing, Swinburne University of Technology, 1 Alfred St, Hawthorn, VIC 3122, Australia\\
$^{12}$ARC Centre of Excellence for Dark Matter Particle Physics (CDM), Australia\\
$^{13}$ARC Centre of Excellence for All-Sky Astrophysics in 3 Dimensions (ASTRO 3D), Australia}

\date{Accepted 2024 June 21. Received 2024 June 20; in original form 2023 September 30}

\pubyear{2024}

\begin{document}
\label{firstpage}
\pagerange{\pageref{firstpage}--\pageref{lastpage}}
\maketitle

\begin{abstract}
Numerous metrics exist to quantify the dynamical state of galaxy clusters, both observationally and within simulations. Many of these correlate strongly with one another, but it is not clear whether all of these measures probe the same intrinsic properties. In this work, we use two different statistical approaches -- principal component analysis (PCA) and uniform manifold approximation and projection (UMAP) -- to investigate which dynamical properties of a cluster are in fact the best descriptors of its dynamical state. We use measurements taken directly from \threehun\ suite of galaxy cluster simulations, as well as morphological properties calculated using mock X-ray and SZ maps of the same simulated clusters. We find that four descriptions of dynamical state naturally arise, and although correlations exist between these, a given cluster can be ``dynamically relaxed'' according to all, none, or some of these four descriptions. These results demonstrate that it is highly important for future observational and theoretical studies to consider in which sense clusters are dynamically relaxed. Cluster dynamical states are complex and multi-dimensional, and so it is not meaningful to classify them simply as ``relaxed'' and ``unrelaxed'' based on a single linear scale.
\end{abstract}

\begin{keywords}
galaxies: clusters: general -- methods: numerical -- dark matter -- galaxies: kinematics and dynamics -- X-rays: galaxies: clusters
\end{keywords}



\section{Introduction}
\label{sec:intro}

The structure of a galaxy cluster consists of a large dark matter halo, typically with a mass between $10^{14}-10^{15}\ M_{\odot}$, filled with hot intracluster gas, and populated with hundreds or thousands of galaxies. A wide range of physical processes take place within clusters, meaning they play a crucial role in many areas of astrophysics and cosmology.

The evolution of galaxies is strongly dependent on their cosmic environment, and clusters represent one of the most extreme environments for a galaxy. Strong tidal forces due to a cluster's gravitational potential can disrupt the structure of a galaxy, dramatically changing its morphology \citep{moore1996, mihos2017, knebe2020,haggar2023}. Additionally, processes such as ram-pressure stripping \citep{gunn1972} by the intracluster gas can remove both the cold gas in a galactic disk and the surrounding halo gas, quenching star formation in galaxies \citep{larson1980, zabel2019}. Galaxy clusters are also an important tool in constraining cosmology -- for example, studying the shapes of cluster halo density profiles can provide information about the nature of dark matter \citep{eckert2022, limousin2022}. Galaxy clusters can also be used as a proxy for measuring cosmological parameters, either through cluster number counts from large surveys \citep[for example]{evrard1989, dehaan2016, abdullah2020}, or by using cluster properties. For example, \citet{amoura2021} measure the formation times of clusters in a suite of simulations with varying values of $\Omega_{\rm{M}}$, the matter density parameter, and $\sigma_{8}$, the amplitude of the matter power spectrum at a scale of \mbox{$8\ h^{-1}\rm{Mpc}$}, and show that the formation times of galaxy clusters depend on the values of $\Omega_{\rm{M}}$ and $\sigma_{8}$.

One of the core properties of a galaxy cluster is its dynamical state -- that is, whether or not it is dynamically relaxed. Numerous areas of galaxy cluster physics, including many of those described above, are dependent on an understanding of cluster dynamical states. It has been shown that various phenomena in the intra-cluster medium (ICM), such as turbulence, differ between relaxed and unrelaxed clusters \citep{lau2009, vallezperez2021, simonte2022}. Furthermore, \citet{nagai2011} and \citet{vazza2013} both show that, in dynamically disturbed clusters, the hot X-ray-emitting ICM contains dense, cooler clumps of gas, associated with infalling galaxies and galaxy groups. The intra-cluster medium drives galaxy evolution mechanisms such as ram-pressure stripping; this has also been shown to be enhanced in dynamically disturbed groups and clusters \citep{mauduit2007, roberts2017, lourenco2023}.

Additionally, many of the astronomical and cosmological measurements described above are indirectly related to the cluster dynamical states, as they rely on accurate measurements of the masses of clusters, which are in turn dependent on a their dynamical state. Cluster masses are often calculated under the assumption that a cluster is in hydrostatic equilibrium. As a result of this, clusters that are dynamically unrelaxed (and so are not in hydrostatic equilibrium) can have their masses underestimated by up to $20\%$ \citep{nagai2007, kravtsov2012}. Relaxed clusters can also have their masses underestimated, albeit to a lesser extent \citep{lau2013, gianfagna2021}, and the strength of this bias is dependent on the redshift of the cluster in question \citep{bennett2022}. Cluster scaling relations also differ between clusters that are relaxed, and those that are rapidly accreting material \citep{planelles2009, lau2015, chen2019}. Additionally, properties such as location of a cluster's splashback radius \citep{adhikari2014, more2015} are dependent on its dynamical state. This is the radius within which the cluster material dominates over the surrounding infalling material, and so this implies that the region in which a galaxy is impacted by a cluster is also dependent on cluster dynamical state. Because of this, a thorough understanding of cluster dynamical states is vital if we are to use clusters as an astronomical and cosmological tool.

In its simplest form, a system of collisionless particles can be described as dynamically relaxed and virialised once the velocities of particles in the system are uncorrelated with their initial velocities. Equivalently, this means the average magnitude of the velocity of each particle, $\bm{v}$, is equal to the change in velocity of the particle due to interactions with other particles, $\delta{\bm{v}}$. The typical time required for a system to reach this stage is given by the relaxation time, $t_{\rm{relax}}$,
\begin{equation}
    t_{\rm{relax}}\approx\frac{0.1N}{{\rm{ln}}(N)}t_{\rm{cross}}\,,
    \label{eq:relaxation_time}
\end{equation}
where $N$ is the number of particles in a system, and $t_{\rm{cross}}$ is the average crossing time for a particle in the system \citep{binney1987}. For a galaxy cluster, the number of particles (galaxies) is $\sim10^{3}$, and the crossing time is \mbox{$\sim1\ \rm{Gyr}$}, leading to a typical relaxation time of \mbox{$\sim10\ \rm{Gyr}$}, comparable to the age of the Universe. However, in the context of galaxy clusters, this definition of dynamical state is not particularly useful for several reasons. Firstly, clusters are not collisionless systems. Galaxies in clusters frequently experience near-misses or tidal interactions \citep{knebe2006, muldrew2011, bahe2019}, and the intra-cluster gas, which makes up a significant portion of a cluster's mass, is not a collisionless fluid. Secondly, this definition describes a closed system, which galaxy clusters are not; clusters are continuously accreting material from their surrounding environment. As such, the effects from their ``boundary'' need to be included to quantify their dynamical state. Finally, according to this definition, only material that has been in a cluster for greater than $10\ \rm{Gyr}$ can be dynamically relaxed. This means that a $z=0$ cluster can only be truly relaxed if it accreted all, or most, of its material before $z=1.5$. While this is technically possible, it is an exceptionally rare scenario in a typical $\Lambda$CDM cosmology.

Consequently, throughout the literature, numerous properties of clusters are treated as indicators of ``dynamical state'', each of which is used to quantify how relaxed is a galaxy cluster. Many of these are theory-based metrics, taken from simulations of galaxy groups or clusters. In \citet{cui2017}, the dynamical state of a cluster is described by three properties: dynamically unrelaxed clusters are those with a centre of mass that is offset from the cluster halo density peak, those with large amounts of substructure, and those that are not virialised. 

This combination of observables has been widely used for some time. For example, \citet{neto2007} classify dark matter haloes based on these three parameters, and place a limit on each of these, describing relaxed haloes as those that satisfy all three of these conditions, and unrelaxed haloes as those that do not. \citet{cui2018} place similar constraints on simulated galaxy clusters, and use this to classify clusters as relaxed and unrelaxed. \citet{haggar2020}, by contrast, combine the three into a single parameter, $\chi_{\rm{DS}}$, giving a single continuous parameter describing how relaxed is a cluster. Other work also uses these measures, such as \citet{wen2013}, who use the amount of substructure as an indicator of dynamical state. They describe how a large amount of structure is indicative of a recent merger event. The results of \citet{kimmig2023} also demonstrate this -- they show that the amount of substructure, and in particular the size of the eighth most massive substructure, are indicative of the merger history of a cluster over the last two gigayears. Further works have shown that numerous other cluster properties correlate with recent merger activity, such as the virialisation and centre of mass offset of a cluster \citep{power2012}, its concentration \citep{wang2020}, or similarly its sparsity \citep{richardson2022}. \citet{contrerassantos2022} also use the time since a major merger as an measure of cluster relaxation, and show how this correlates with the $\chi_{\rm{DS}}$ measure from \citet{haggar2020}. 

Already, it is apparent that different measures of a cluster's dynamical state are probing different core properties. Measurements such as the substructure fraction and centre of mass offset are quantities that can be measured observationally at a single epoch, although in observations they are projected into two dimensions. This is in contrast to the time since the last major merger, which is a property of the history of a cluster. Other studies also take this approach to dynamical state, connecting it to the total history of a cluster. For example, \citet{diemer2014} introduced the accretion rate proxy $\Gamma$, variants of which have since become widely used as measures of halo relaxation \citep[e.g.][]{vallezperez2020}. An alternative approach is to use a redshift-dependent definition of dynamical state, based on the fact that different cluster properties evolve over different timescales \citep[e.g.][]{mendoza2023, vallezperez2023}.

\citet{gouin2021} probe the dynamical state of clusters from the IllustrisTNG simulation \citep{nelson2019} using three separate measures of cluster growth history: the $z=0$ halo growth rate, the $z=0.5$ mass accretion rate, and the cluster formation time ($z_{0.5}$, see \Sec{sec:3d_measures}). They also show that dynamically disturbed clusters are more strongly connected to the cosmic web -- that is, they have more cosmic filaments attached to the cluster \citep[see][for a similar study with \tth\ clusters]{santoni2024}. Similarly, \citet{darraghford2019} find that simulated clusters that have recently experienced a major merger have a higher connectivity. This is yet another interpretation of dynamical state, as a property of the surrounding region of the Universe. This cosmic environment can also impact global properties of the cluster and its dark matter halo, such as its shape \citep{gouin2021, smith2023} and concentration \citep{neto2007}.

Additional complexity comes from the large number of cluster dynamical state metrics and morphological parameters that are used in observational astronomy. For example, cluster shapes can be mapped using X-ray and Sunyaev-Zel'dovich (SZ) effect observations. From these, many quantifiable properties can be measured, such as the amount of substructure \citep{ge2016}, offset of the brightest cluster galaxy (BCG) from the X-ray/SZ peak \citep{zenteno2020} or the power spectrum of the hot gas distribution \citep{cerini2023}. Combinations of multiple X-ray or SZ morphological parameters have been shown to provide even more robust measures of dynamical state than these individual parameters (\citet{parekh2015, yuan2020, campitiello2022, yuan2022}; see \citet{rasia2013} and \citet{zhang2024} for more comprehensive discussions). Other observational studies include \citet{wen2013}, who quantify cluster substructure using optical data from the Sloan Digital Sky Survey \citep[SDSS;][]{aihara2011}. 

The result of having so many different properties in common use is that the ``dynamical state'' of a cluster is not clearly or consistently defined in the literature. Consequently, when the dynamical states of clusters are inferred from observations, it is not entirely apparent which fundamental property of a cluster these are probing. It is also not clear if they are all probing the same intrinsic cluster property, or if the ``dynamical state'' of a cluster is actually made up of several properties.

In this study, we aim to gain a deeper understanding of cluster dynamical state measurements by investigating the connections between theory-based and observable properties of dynamical state, the degeneracy between different measures of dynamical state, and the core, intrinsic properties of clusters that these measurements are actually probing. Our primary approach to this is through dimensionality reduction -- reducing a large number of dynamical state metrics to a smaller set of variables will make the nature of dynamical state clearer and easier to interpret. Previous studies have taken a similar approach to ours, although mostly to consider the accretion histories of dark matter haloes, rather than the dynamical states of galaxy clusters. One notable example is \citet{wong2012}, who use principal component analysis \citep[PCA;][]{jolliffe2016} to determine the principal components of 10 input properties of dark matter haloes, and study the mass accretion histories of these haloes. They show that splitting these clusters by their first and second principal components naturally displays two separate modes of accretion history: the halo formation time, and the acceleration/deceleration of a halo's accretion. For further discussion, we also refer the reader to \citet{jeesondaniel2011} and \citet{skibba2011}, who perform similar analyses on simulations of dark matter haloes.

In this work we develop the methods used in previous studies, applying principal component analysis to 17 indicators of cluster dynamical state, based on 3D data (as opposed to directly observable quantities) from simulations of $z=0$ galaxy clusters. This analysis uses \threehun\ project, a suite of hydrodynamical zoom simulations of massive galaxy clusters, taken from a large \mbox{$1\ h^{-1}\rm{Gpc}$} cosmological volume. We physically interpret the principal components that naturally arise from this analysis, and show how they correspond to the mass accretion histories of these clusters. Despite this work focusing on a theoretical approach to dynamical states, we also compare our PCA to X-ray and SZ properties measured from mock observations of the same simulated clusters, originally calculated in \citet{deluca2021}. Finally, we expand further on previous studies in this area by analysing the same simulated clusters using uniform manifold approximation and projection \citep[UMAP;][]{mcinnes2018}, an alternative approach to dimensionality reduction. While we do not develop a quantitative means to classify clusters, we demonstrate qualitatively the different dynamical states of clusters that exist in our simulations.

We interpret our results as showing that the dynamical state of a cluster is not a single property. Instead, a single cluster has multiple ``dynamical states'', and can be relaxed in all of these dimensions, or none of them, or \textit{some} of them.

The paper is structured as follows: in \Sec{sec:methods} we introduce the simulation data used throughout this work, as well as the dynamical state indicators we use directly from the simulations (\Sec{sec:3d_measures}) and from mock X-ray and SZ-effect maps generated from the simulation data (\Sec{sec:morph_parameters}). In \Sec{sec:results} we study these parameters using both PCA and UMAP, and discuss how the results connect to the mock observations of clusters, and their mass accretion histories. Finally, in \Sec{sec:conclusions} we summarise our findings, and the implications this has for how the dynamical state of galaxy clusters should be interpreted.

\section{Simulations and methodology}
\label{sec:methods}

This work utilises simulation data from \threehun\ project. We study these galaxy clusters in the final snapshot of the simulations ($z=0$). In \Sec{sec:simulation_data} we provide an overview of the simulation data, but for a detailed description we refer the reader to \citet{cui2018}.

\subsection{Simulation data}
\label{sec:simulation_data}

\threehun\ project (hereafter \tth) is a suite of hydrodynamical resimulations of large galaxy clusters. The simulations are based on the MDPL2 MultiDark simulation \citep{klypin2016}\footnote{The MultiDark simulations are publicly available from the CosmoSim database, \url{https://www.cosmosim.org}.}. MDPL2 is a large dark matter-only simulation, with a comoving box size of \mbox{$1\ h^{-1}\rm{Gpc}$}, which uses {\textit{Planck}} cosmology ($\Omega_{\rm{M}}=0.307$, $\Omega_{\rm{B}}=0.048$, $\Omega_{\Lambda}=0.693$, $h=0.678$, $\sigma_{8}=0.823$, $n_{\rm{s}}=0.96$) \citep{planck2016}\footnote{The reduced Hubble constant, $h$, is defined such that the Hubble constant, \mbox{$H_{\rm{0}}=h\times100\ \rm{km}\ \rm{s}^{-1}\ \rm{Mpc}^{-1}$}.}.

From this simulation, the 324 most massive dark matter haloes were selected, and resimulated from their initial conditions (at an initial redshift of $z_{\rm{init}}=120$) with baryonic physics included. For each cluster, the \mbox{$1\ h^{-1}\rm{Gpc}$} dark matter-only simulation box was re-centred on the cluster, and the particles within \mbox{$15\ h^{-1}\rm{Mpc}$} of the cluster centre at \mbox{$z=0$} were traced back to their initial conditions. Each of these particles was then split into a dark matter particle and a gas particle, with masses of \mbox{$m_{\rm{DM}}=1.27\times10^{9}\ h^{-1}M_{\odot}$} and \mbox{$m_{\rm{gas}}=2.36\times10^{8}\ h^{-1}M_{\odot}$} respectively, set according to the cosmic baryon fraction. The resolution of dark matter particles outside of this radius was degraded, allowing the large-scale tidal forces acting on the cluster to be maintained with a reduced computational cost. 

In this work, we use clusters from \tth\ dataset simulated using the {\sc{GadgetX}} code. {\sc{GadgetX}} is an updated version of the {\sc{Gadget3}} code \citep{springel2005, beck2016, sembolini2016}, and uses a smoothed-particle hydrodynamics scheme to simultaneously evolve the baryonic and dark matter components of a simulation. As well as gas, the stochastic star formation in {\sc{GadgetX}} produces stellar particles of varying mass, typically around \mbox{$m_{\rm{star}}\sim4\times10^{7}\ h^{-1}M_{\odot}$} \citep{tornatore2007, murante2010, rasia2015}. Type II supernova feedback and AGN feedback are included, as described in \citet{springel2003} and \citet{steinborn2015} respectively. The final dataset consists of 324 galaxy clusters, with masses ranging from \mbox{$M_{200}=5\times10^{14}\ h^{-1}M_{\odot}$} to \mbox{$M_{200}=2.6\times10^{15}\ h^{-1}M_{\odot}$}. Here, $M_{200}$ is the mass enclosed within a sphere of radius $R_{200}$, defined such that the average density within this sphere is equal to 200 times the critical density of the Universe at that redshift. For the cluster masses in \tth, the \mbox{$15\ h^{-1}\rm{Mpc}$} high-resolution region corresponds approximately to \mbox{$7-10R_{200}$}. As well as the extensive description in \citet{cui2018}, further description of \tth\ dataset is available in other previous studies that have used these data. We particularly refer the reader to studies that have investigated cluster dynamical states using \tth\ \citep[for example]{haggar2020, capalbo2021, deluca2021, contrerassantos2022, li2022}.

In this work, we specifically use the data from the final snapshot of \tth\ simulations ($z=0$), as the focus of this work is on present-day galaxy clusters. However, some of the cluster properties we calculate also rely on cluster properties at $z>0$ (see \Sec{sec:3d_measures} for details). This information requires the construction of halo merger trees, which are described in the following section.

\subsubsection{Galaxy identification and tree-building}
\label{sec:galaxy_identification}

In each snapshot of \tth\ data, the haloes and subhaloes are identified using the Amiga Halo Finder, {\sc{ahf}}\footnote{\url{http://popia.ft.uam.es/AHF}} \citep[see][for a detailed description of {\sc{ahf}}]{gill2004_ahf, knollmann2009}. {\sc{ahf}} is a density peak halo finder, and is used to identify the particles in the main cluster halo, subhaloes of the cluster, and haloes in the surrounding region. For each of these (sub)haloes, properties such as the position, velocity and mass are given as outputs. The halo catalogues in each snapshot were linked together using the {\sc{mergertree}} tree-builder, which is part of the {\sc{ahf}} package. This tree-builder uses a merit function, $M_{\rm{i}}$, given in Table B1 of \citet{knebe2013}, to identify the main progenitor of each halo by searching for particles that are common between the two haloes. {\sc{mergertree}} is also able to ``skip'' snapshots, meaning that if {\sc{ahf}} is unable to resolve a halo in one snapshot, the tree-builder can instead find an appropriate progenitor in an earlier snapshot \citep[see][for additional details on {\sc{mergertree}}]{knebe2011_halo_comparison, srisawat2013}.

For most of this work, we include all of these subhaloes (and thus all of the particles) in the calculations of dynamical state parameters. For those measures that rely on galaxy properties (e.g. cluster richness, $N_{200}$, and magnitude difference between galaxies, $m_{12}$, detailed in \Sec{sec:3d_measures}), we include all galaxies with a total mass greater than \mbox{$10^{10.5}\ h^{-1}M_{\odot}$}, and a stellar mass of greater than \mbox{$10^{9.5}\ h^{-1}M_{\odot}$}. These limits have an extremely minor impact on our results, as most dynamical state indicators based on galaxy properties are dependent on the largest, brightest galaxies in a cluster.

\subsection{Measures of dynamical state}
\label{sec:dynamical_state}

\subsubsection{3D measures}
\label{sec:3d_measures}

From the simulations, we utilise 17 different properties, each of which is associated with cluster dynamical state. These are detailed below, along with relevant information on how they were calculated in our simulations. Many of these quantities are calculated using $R_{200}$ as an outer boundary, as this is the characteristic radius used by {\sc{ahf}} in calculating halo properties; we state explicitly when this is not the case. These properties are also calculated using all particles within the relevant radius (typically $R_{200}$ unless stated otherwise), apart from $c_{200}$ which only uses dark matter particles.

\begin{itemize}
\item $f_{\rm{s}}$: The fraction of cluster mass inside a given radius that is contained within subhaloes. Two values of this are used, $f_{\rm{s}}(R_{200})$ and $f_{\rm{s}}(R_{500})$, equal to the substructure mass fraction inside the radii $R_{200}$ and $R_{500}$ respectively. Using these two radii allows us to probe both the outer and inner regions of a cluster: as \citet{cui2017} show, dynamical state indicators are dependent on cluster-centric distance.
\item $\Delta_{\rm{r}}$: The offset of the centre of mass of the cluster from the density peak of the cluster halo, as a fraction of the cluster radius $R_{200}$. Similarly to $f_{\rm{s}}$, $\Delta_{\rm{r}}(R_{200})$ and $\Delta_{\rm{r}}(R_{500})$ are both used, which are each calculated using the centre of mass of all material inside the radii $R_{200}$ and $R_{500}$ respectively.
\item $\eta$: The virial ratio, a measure of how well a cluster obeys the virial theorem, based on its total kinetic energy, $T$, its total potential energy, $W$, and its energy from surface pressure, $E_{\rm{s}}$. It is typically defined in the literature as $\tilde{\eta}=(2T-E_{\rm{s}})/|W|$, so that $\tilde{\eta}=1$ for virialised haloes. However, PCA is designed to capture linear, monotonic relationships between variables. This variable, where the ``extreme'' cases (most virialised haloes) correspond to an intermediate value ($\tilde{\eta}=1$) is therefore not well-suited to PCA. Consequently, we perform a transformation, defining a new $\eta$ such that:

\begin{equation}
\eta=\left|\frac{\left(2T-E_{\rm{s}}\right)}{|W|}-1\right|\,.
\label{eq:eta}
\end{equation}
Increasing values of this quantity correspond to a greater deviation from virialisation, and thus to less virialised haloes, making this quantity better suited for use in PCA. Note also that this virial ratio differs from the classic definition of virialisation, due to the additional surface pressure term which accounts for clusters' ongoing accretion of material \citep{poole2006, shaw2006}. The surface pressure is calculated as the energy from surface pressure integrated over the halo boundary -- a detailed mathematical description is given in Section 3 of \citet{cui2017}. The virial ratio is also calculated twice, for all material inside $R_{200}$ and $R_{500}$.
\item $z_{0.5}$: The formation time of a cluster: the redshift at which the cluster mass, $M_{200}$, is equal to half its value at $z=0$.
\item $\lambda$: A dimensionless spin parameter, used to describe the bulk rotation of a cluster. It is defined in the same way as \citet{bullock2001},
\begin{equation}
    \lambda=\frac{J_{200}}{\sqrt{GM_{200}^{3}R_{200}}}\,,
    \label{eq:bullock_spin}
\end{equation}
where $J_{200}$ is the total angular momentum of material inside $R_{200}$. We note that other definitions of the cluster spin also exist \citep[e.g.][]{peebles1969}, but we use this value as it is calculated using only mass within a well-defined radius, $R_{200}$.
\item $c/a$: Sphericity, the ratio of the minor and major axes of the cluster halo's moment of inertia tensor. As this calculation of the moment of inertia includes all particles (dark matter, gas, and stars), it accounts for these particles' varying masses accordingly. We note that several alternative measures of cluster shape are also used throughout the literature, including triaxiality, ellipticity, and prolaticity \citep{lau2021}, but we only use sphericity in this work. 
\item $c_{\rm{200}}$: The concentration of the dark matter halo. This is equal to the ratio between $R_{200}$ and $R_{\rm{s}}$, the scale radius of a halo, as defined by an NFW profile \citep{navarro1996}. Here, $\rho(r)$ is the dark matter density of a halo as a function of halo-centric distance, and $\rho_{0}$ is some characteristic density:
\begin{equation}
    \rho(r)=\frac{\rho_{0}}{\left(\frac{r}{R_{\rm{s}}}\right)\left(1+\frac{r}{R_{\rm{s}}}\right)^{2}}\,.
\label{eq:nfw}
\end{equation}
This is an output from {\sc{ahf}}, which selects cluster centres based on a density peak finder. The concentration of a halo is not calculated directly, but is instead a numerical solution to Equation 9 in \citet{prada2012}, based on the maximum circular velocity of a cluster.
\item $z_{\rm{merge,50}}$: The redshift at which a cluster last experienced a merger that increased its mass by more than $50\%$. These are equal to the values of $z_{\rm{start}}$ calculated in \citet{contrerassantos2022}, which describe the jump in the mass accretion history of the cluster and therefore the onset of a merger phase.
\item $\gamma$: The mass accretion rate of the cluster, defined as the fractional increase in $M_{200}$ within the last dynamical time, $t_{\rm{d}}$. We use the crossing time of a cluster as its dynamical time \citep[see][for example]{binney1987,contrerassantos2022}, equal to:
\begin{equation}
    t_{\rm{d}}\simeq\frac{R_{200}}{v_{\rm{circ}}}=\sqrt{\frac{R_{200}^{3}}{GM_{200}}}\,.
\label{eq:dynamical_time}
\end{equation}

Using the definition of $M_{200}$, and the fact that the critical density $\rho_{\rm{crit}}=3H^{2}/8\pi G$, we find that the dynamical time at $z=0$ is given by $t_{\rm{d}}\simeq1/(10H_{0})$. For the cosmology used in \tth\ simulations, this dynamical time is equal to approximately $1.4$ Gyr, or a redshift of $z=0.1$. Hence, the value of $\gamma$ is the fractional increase in mass between $z=0.1$ and $z=0$.
\item $N_{\rm{fil}}$: The number of filaments, or connectivity, of a cluster. The connectivity of each cluster was calculated using the DIScrete PERsistent Structure Extractor (DisPerSE) filament finding algorithm \citep{sousbie2011}. In our specific case, the cosmic filaments were identified based on the number density of gas particles around a cluster. Multiple definitions for the connectivity of clusters exist -- we use it to refer to the number of these filaments beginning at the cluster centre (node) and passing through a sphere of radius $R_{200}$ surrounding the cluster. Cosmic filaments around the clusters in \tth\ have been studied extensively in other works via galaxies \citep{cornwell2022, kuchner2022} and their gas component \citep{santoni2024}.
\item $D_{n,f}$: This environment parameter is defined as the distance to the $n$th nearest halo whose mass is greater than $fM_{200}$, in units of $R_{200}$. In our case, we use $n=1$, $f=0.1$, and so $D_{1,0.1}$ is the distance to the nearest halo whose mass is greater than one tenth of the cluster's mass \citep[see also][]{jeesondaniel2011,wong2012}. This is consequently a measure of how isolated a galaxy cluster is from other clusters with masses of the same order of magnitude.
\item $N_{200}$: Richness, the number of cluster members whose absolute magnitude is between $m_{3}$ and $m_{3}+2$, where $m_{3}$ is the magnitude of the third-brightest cluster member \citep[this definition is given by][and while multiple definitions of cluster richness exist, this is the one we use throughout this work]{abell1958}. We use $R$-band luminosities, calculated using the stellar population synthesis code STARDUST \citep[see][]{devriendt1999}.
\item $m_{12}$: Difference in magnitude between the brightest and second-brightest cluster member galaxies, or ``magnitude gap''. Multiple names are used throughout the literature to refer to this property, or similar properties \citep[e.g. the ``fossil parameter'' in ][also used as a measure of dynamical state]{ragagnin2019}. As in our calculations of cluster richness, we calculate this using $R$-band magnitudes.
\item $\sigma_{\rm{BCG}}$: The velocity dispersion of stars in the brightest cluster galaxy (BCG) of our simulated clusters. This was calculated using the stellar particles located within a spherical aperture of radius \mbox{$200\ h^{-1}{\rm{kpc}}$} around the centre of the cluster halo, a radius which should include most of the material associated with the BCG \citep{lin2004,contrerassantos2022}.
\end{itemize}

It should be noted that the order in which we present these 17 dynamical state indicators is arbitrary. In principle, one could group these parameters together -- for example, $z_{0.5}$, $z_{\rm{merge,50}}$ and $\gamma$ all depend on temporal information about a cluster. However, we have not chosen to do this here, as these properties are taken by our analysis in a random order, without any further information relating them to one another. Later, we will deliberately reorder these 17 parameters based on our principal component analysis (see \Sec{sec:pca}).

These measurements are all dependent on information that is only available from the full 3D version of our simulations. While the strength of this dependence varies, it means that none of these are directly measurable from observations. Some of these properties can be measured quite well in observations, such as the velocity dispersion of the BCG, $\sigma_{\rm{BCG}}$, which can be measured using spectroscopy, and the magnitude difference between the two brightest galaxies, $m_{12}$. Conversely, some are much more sensitive to 2D projection effects. For example, the centre of mass offset, $\Delta_{\rm{r}}$, is strongly dependent on the viewing angle, particularly if the offset is caused by a single major merger event. If the merger takes place along the line of sight, the apparent offset will be minimal. If instead it occurs in the plane of the sky, the calculated offset of the centre of mass will be apparent and measurable \citep[see][]{zenteno2020}. A similar property is the sphericity, $c/a$ -- very different values of this will be measured depending on whether a cluster's major axis is aligned along the line of sight or not. Other properties such as the substructure fraction, $f_{\rm{s}}$, and cluster richness, $N_{200}$, are likely to be measured quite well in observations, but will still suffer somewhat due to the presence of interloper galaxies along the line of sight. Additionally, some of these cluster properties cannot be measured at all. The cluster formation time, $z_{0.5}$, requires knowledge of the growth histories of a cluster over several gigayears, which can be inferred from other properties but not measured directly.

The fact that some of these properties are difficult, or impossible, to measure observationally is a topic we plan to address in a future study (see also \Sec{sec:conclusions}). However, in this work our focus is on the actual properties of a cluster, rather than the limitations of what can be measured. The exception is the mock X-ray and SZ data that we use, which we describe in the following subsection.

\subsubsection{X-ray and SZ morphological parameters}
\label{sec:morph_parameters}

In \citet{deluca2021}, mock observations were created for all the clusters of \tth\ sample. Maps were generated for each of these clusters in X-rays, and also as they would be seen through the SZ effect, in terms of the Compton parameter. From these maps, six morphological parameters are calculated, as described below. More rigorous descriptions of these parameters, as well as details on the production of the mock maps and a more thorough review of the literature, can be found in \citet{deluca2021}.

\begin{itemize}
\item $A$: Asymmetry \citep{schade1995}, the normalised difference in flux between the original map, and a rotated map. The value of $A$ chosen is the maximum calculated from four different rotations/reflections ($90\degree$, $180\degree$, and reflection along the main cluster axes).
\item $K$: Light concentration ratio \citep{santos2008}, the ratio of surface brightness computed within two concentric apertures. For the X-ray maps these are $0.025R_{500}$ and $0.25R_{500}$, and for the SZ maps these radii are $0.05R_{500}$ and $0.25R_{500}$.
\item $W$: Centroid shift \citep{mohr1993}, a measure of how much the centroid of a map shifts as different apertures are used to calculate the centre.
\item $P$: Power ratio \citep{buote1995}, is based on a multipole decomposition applied to the maps of the ICM. Specifically, $P$ is the third order power ratio.
\item $G$: Gaussianity \citep{cialone2018}, the ratio of the two values for standard deviation required to describe a 2D Gaussian fit to the map. This can distinguish elongated and circular clusters, and so is analogous to the cluster sphericity, $c/a$.
\item $S$: Strip variation \citep{cialone2018}, the normalised difference between four light profiles, inclined by $45\degree$ to one another, passing through the centroid.
\end{itemize}

Additionally, \citet{deluca2021} calculate $M$, a normalised, linear sum of these six morphological parameters, each weighted such that the difference in $M$ between relaxed and unrelaxed clusters is maximised. Each of these seven total parameters (six parameters, plus the combined measure) is calculated for both X-ray and SZ-effect maps, and we use subscripts to distinguish between these. For an additional study using similar methods, we also refer the reader to \citet{campitiello2022}.

\section{Results}
\label{sec:results}

\subsection{PCA of dynamical state indicators}
\label{sec:pca}

Principal component analysis is a commonly used dimensionality reduction technique, which defines new variables (``principal components'') in a multi-dimensional parameter space. These principal components are linear sums of the input parameters, defined by the eigenvectors of the covariance matrix of the data, meaning they are orthogonal and uncorrelated to one another. The principal components can also be ordered based on the variance in the data for which they account, thus allowing one to only consider the ``most important'' components. These components can be interpreted physically, and can be used to identify correlations, trends and degeneracies in high-dimensional data. PCA requires the input data to be standardised; we have applied this to each of our 17 parameters, and also for our subsequent UMAP analysis (\Sec{sec:umap}). Some discussion of the non-standardised distributions can be found in \App{sec:ds_correlations}. 

\begin{figure*}
\includegraphics[width=\textwidth]{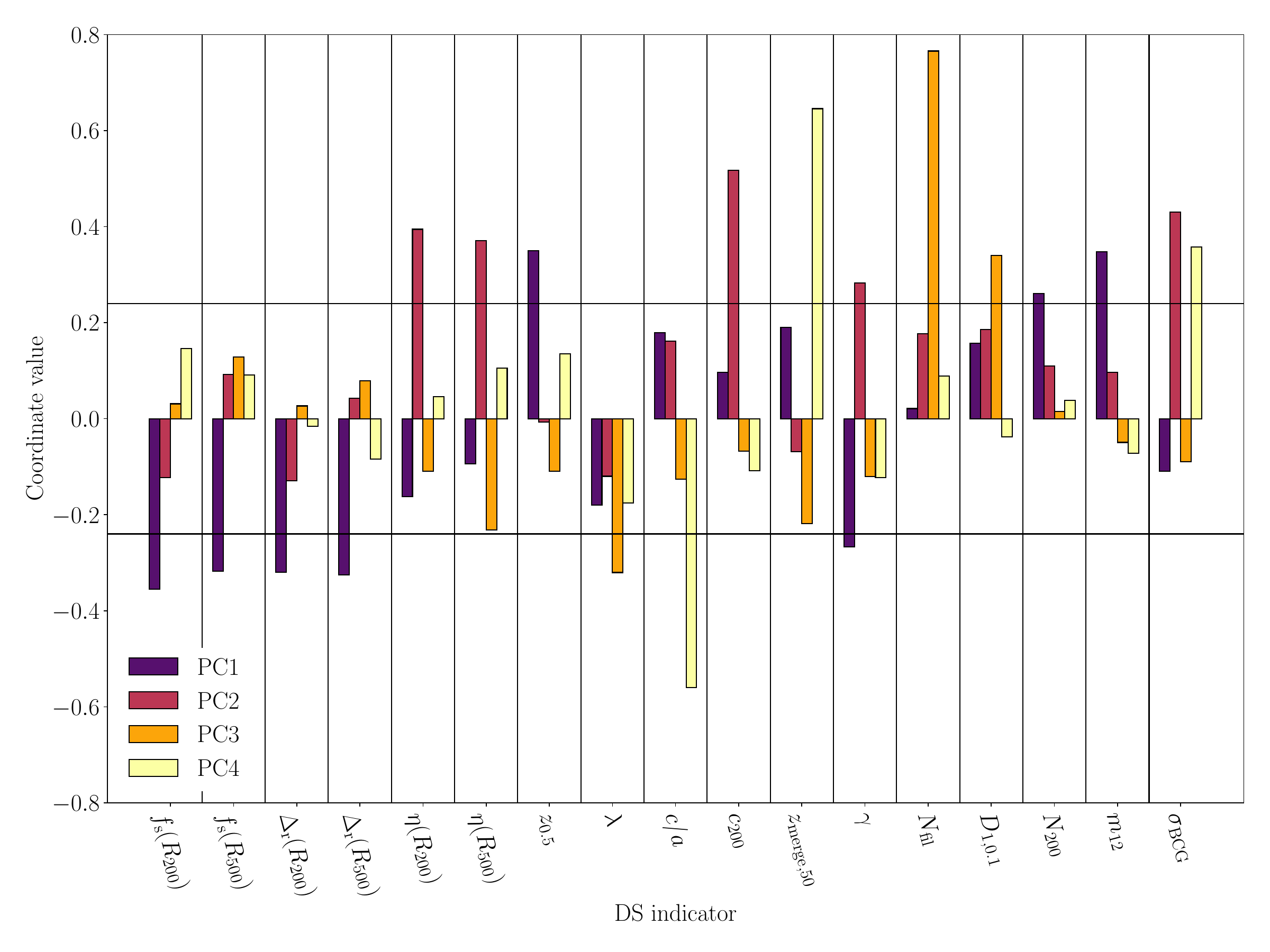}
\caption{Contribution of each of the 17 dynamical state parameters to the four main principal components, as determined by principal component analysis. Dynamical state indicators are listed horizontally, and the value of the bars show the coordinate value of that dimension for PC1, PC2, PC3 and PC4, coloured based on the component number. Horizontal lines are marked at $\pm0.24$, the root-mean-square contribution of any parameter to a given principal component -- this is the boundary at which we consider a parameter to be an ``important'' contributor to that principal component. This data is tabulated in \Tab{tab:prin_comps}.}
\label{fig:pca_bars}
\end{figure*}

\Fig{fig:pca_bars} shows the contributions of each of the 17 dynamical state indicators to the four major principal components, PC1, PC2, PC3 and PC4, as determined by principal component analysis. For each dynamical state indicator, its contribution (i.e. the coordinate value in that dimension) to each of these four components is shown by the height of the bars, coloured based on the component number. Horizontal lines are marked at $\pm0.24$. We take component coordinates that have an absolute value of greater than 0.24 to be ``important'' contributing parameters to a principal component; as these principal components are normalised, the root-mean-square contribution is $N_{\rm{par}}^{-0.5}\approx0.24$, where $N_{\rm{par}}$ is the dimensionality of the data (17 in our case). While this distinction is still somewhat arbitrary, it is in line with previous similar work, such as \citet{wong2012}, who use a boundary of 0.3 for $N_{\rm{par}}=13$. These data are also shown in \Tab{tab:prin_comps}, and component coordinates greater than 0.24 are shown in bold.

\begin{table}
	\centering
	\caption{Coordinate values for the four major dynamical state principal components (this data is also shown in \Fig{fig:pca_bars}). Component coordinates with an absolute value greater than 0.24 are highlighted in bold.}
	\label{tab:prin_comps}
	\begin{tabular}{ccccc} 
		\hline
		Parameter & PC1 & PC2 & PC3 & PC4 \\
		\hline
        $f_{\rm{s}}(R_{\rm{200}})$ & \bf{ -0.36} &  -0.12 &   0.03 &   0.15\\
        $f_{\rm{s}}(R_{\rm{500}})$ & \bf{ -0.32} &   0.09 &   0.13 &   0.09\\
        $\Delta_{\rm{r}}(R_{\rm{200}})$ & \bf{ -0.32} &  -0.13 &   0.03 &  -0.02\\
        $\Delta_{\rm{r}}(R_{\rm{500}})$ & \bf{ -0.33} &   0.04 &   0.08 &  -0.08\\
        $\eta(R_{\rm{200}})$ &  -0.16 & \bf{  0.39} &  -0.11 &   0.05\\
        $\eta(R_{\rm{500}})$ &  -0.09 & \bf{  0.37} &  -0.23 &   0.11\\
        $z_{\rm{0.5}}$ & \bf{  0.35} &  -0.01 &  -0.11 &   0.13\\
        $\lambda$ &  -0.18 &  -0.12 & \bf{ -0.32} &  -0.18\\
        $c/a$ &   0.18 &   0.16 &  -0.13 & \bf{ -0.56}\\
        $c_{\rm{200}}$ &   0.10 & \bf{  0.52} &  -0.07 &  -0.11\\
        $z_{\rm{merge,50}}$ &   0.19 &  -0.07 &  -0.22 & \bf{  0.65}\\
        $\gamma$ &  \bf{ -0.27} &   \bf{ 0.28} &  -0.12 &  -0.12\\
        $N_{\rm{fil}}$ &   0.02 &   0.18 & \bf{  0.77} &   0.09\\
        $D_{1,0.1}$ &   0.16 &   0.19 & \bf{  0.34} &  -0.04\\
        $N_{200}$ &   \bf{ 0.26} &   0.11 &   0.01 &   0.04\\
        $m_{12}$ & \bf{  0.35} &   0.10 &  -0.05 &  -0.07\\
        $\sigma_{\rm{BCG}}$ &  -0.11 & \bf{  0.43} &  -0.09 & \bf{  0.36}\\
        \hline
	\end{tabular}
\end{table}

Throughout the remainder of this work, we consider only these four dominant principal components. Between them, these four explain $64\%$ of the variance of the 17 dynamical state indicators ($38\%$, $14\%$, $6\%$ and $6\%$ for PC1-PC4 respectively), as shown in \Fig{fig:explained_variance}. We choose to consider only four parameters as this is the minimum number such that every dynamical state indicator contributes strongly to at least one principal component, and one of the aims of this work is to group these indicators in as simple a way as possible. While there is not a sharp decrease in the importance of components after PC4, we do not believe that including these additional components allows for significantly more scientific interpretation; this is discussed further in \App{sec:explained_variance}.

The major contributors to each principal component can be summarised as:

\begin{itemize}
    \item PC1: Substructure fraction, centre of mass offset, formation time, accretion rate, cluster richness, and dominance of BCG.
    \item PC2: Virial ratio, concentration, accretion rate, and BCG velocity dispersion.
    \item PC3: Cluster spin, connectivity, and distance to nearest large halo.
    \item PC4: Sphericity, time since last major merger, and BCG velocity dispersion.
\end{itemize}

Throughout the remainder of this work, we have reordered our 17 indicators from the order in which they were presented in \Sec{sec:3d_measures}, based on the principal component to which they strongly contribute (see \Fig{fig:obs_spearmans}, for example).

We can interpret these principal components physically, as four different forms of dynamical state.

PC1 describes the formation time of a cluster, and the properties of galaxies in this cluster -- that is, whether a cluster is a recently-forming rich cluster with many bright galaxies, or an old, poor cluster dominated by a single BCG. This component primarily describes the history of substructure accretion by this cluster, and is the ``most important'' principal component, explaining more than one-third of the total variance in the dataset.

PC2 describes the relaxation and virialisation of the dark matter halo of the cluster. It is dependent on whether the halo is virialised or not, and how concentrated it is. In turn, the BCG velocity dispersion is also included in this. Highly concentrated haloes have a greater BCG velocity dispersion as their central dark matter density is greater, while low-concentration haloes (those with more of a central core) have a lower BCG velocity dispersion.

PC3 represents the local environment of a cluster: how connected it is to cosmic filaments, and whether it is in an isolated region of the Universe. The local environment will impact the shear forces on a cluster, potentially explaining the inclusion of spin in this component. 

PC4 includes the information on whether this cluster is a recent merger. Currently-merging clusters are unlikely to be spherical as they will consist of a superposition of two approximately spherical haloes. Moreover, \citet{contrerassantos2022} use \tth\ data to show that BCG properties are strongly impacted by major mergers, explaining the inclusion of BCG velocity dispersion in this component. 

Interestingly, we note that PC4 actually consists of a positive $z_{\rm{merge,50}}$ component and a negative $c/a$ component, implying that the value of this component is greater for elongated clusters, and those that last merged long ago. While this is not necessarily expected, we believe that this behaviour is a consequence of the definition of $z_{\rm{merge,50}}$, which only includes mergers that have already finished -- ongoing mergers will not be counted as a ``recent merger'', even though their measure of $c/a$ will be strongly impacted \citep[this is the same as the ``reduced'' merger sample in][]{contrerassantos2022}. Additionally, we note that both of these parameters also appear in PC1 (albeit with lower contributions than in PC4) and vary in the same direction. PC1 explains more of the total variance than PC4, and so some of the merger behaviour is also encompassed within PC1, as well as in PC4. This is supported by the fact that there is no significant overall correlation between $z_{\rm{merge,50}}$ and $c/a$ (Spearman's rank, $\rho_{\rm{s}}=0.06$, $p=0.30$, as shown later in \Fig{fig:spearmans_3d}).

Similarly, we note that PC2 includes positive contributions from $\eta$, $c_{200}$, $\gamma$ and $\sigma_{\rm{BCG}}$, indicating that a greater value of this component corresponds to clusters that are non-virialised, rapidly accreting, have a disturbed BCG, and are highly concentrated. Mergers are generally known to make cluster haloes less concentrated, and so this result is also unexpected. It can potentially be explained by the fact that some of the components of PC2 (such as $\gamma$) also contribute to PC1, similarly to how some elements of PC4 contribute to PC1. Also, as shown in \citet{wang2020}, the relationship between mergers and halo concentration can be somewhat complex and non-monotonic; we discuss this further in \Sec{sec:umap}.

Each of these four principal components are, by definition, orthogonal and uncorrelated, and they are each driven by specific properties of a cluster. However, it is important to note that these four different forms of the dynamical state of a galaxy cluster are not independent of one another; multiple dynamical state indicators contribute in a non-negligible way to several principal components. We do not perform a detailed, qualitative analysis of the relationship between these different forms in this work, but instead focus on exploring these four different dimensions of dynamical state.

\subsection{X-ray and SZ morphological indicators}
\label{sec:xray_sz_results}

This paper is primarily a theoretical study, focusing on properties that are taken directly from simulations, rather than from mock observations of simulated clusters. In this section we draw some connections to observational astronomy, by comparing these parameters to mock X-ray and SZ observations of the same galaxy clusters. 

\begin{figure*}
\includegraphics[width=\textwidth]{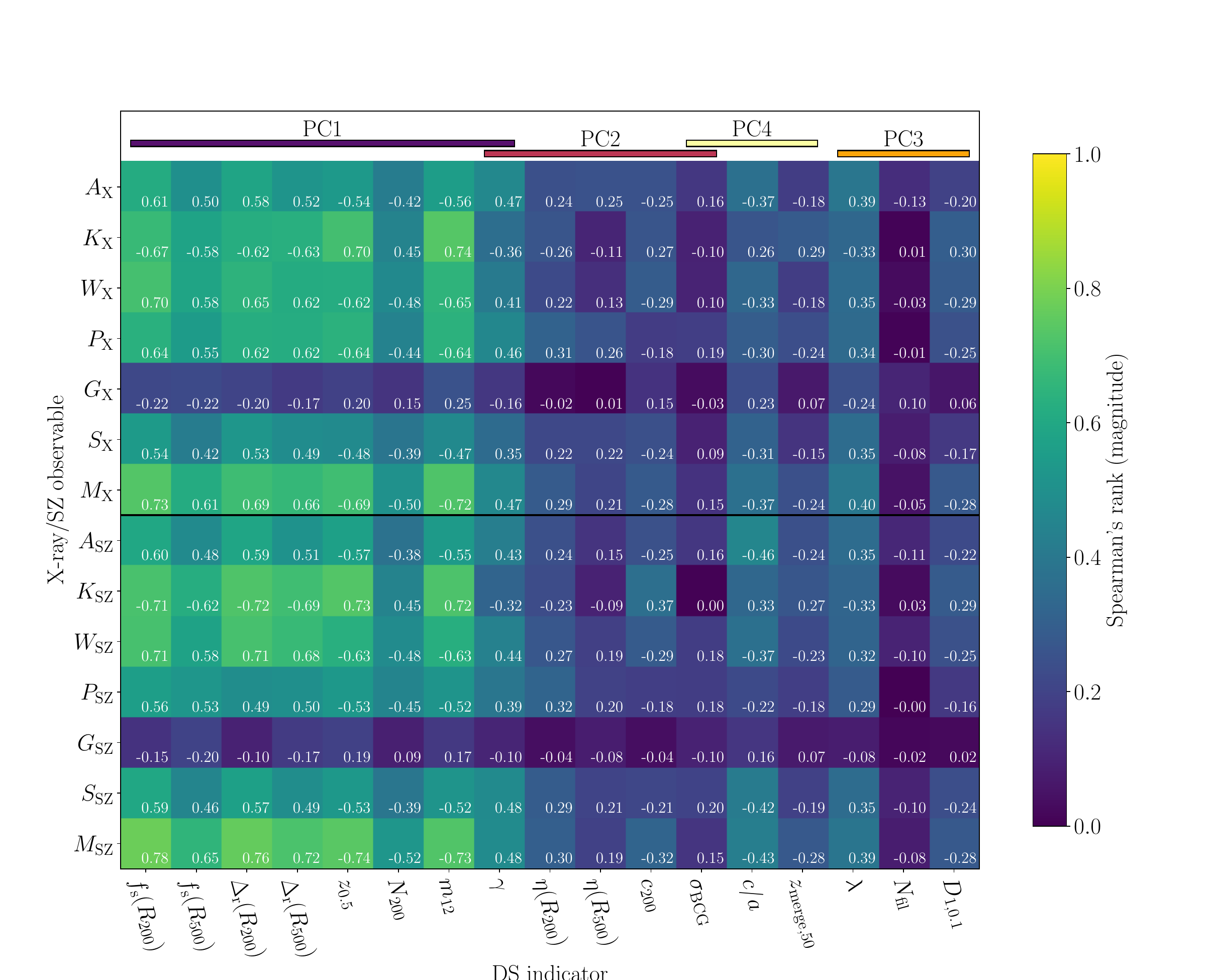}
\caption{Spearman's rank correlation coefficient, $\rho_{\rm{s}}$, between the 17 dynamical state parameters described in \Sec{sec:3d_measures}, and the seven observable morphological parameters described in \Sec{sec:morph_parameters}, for both X-ray and SZ mock observations (separated by a horizontal black line, with X-ray measurements in the top half). The colour of each cell represents the (absolute) value of the Spearman's rank; lighter colours represent a stronger correlation. The value of the Spearman's rank is also written in each cell. Note that the 17 dynamical state indicators have been reordered, and grouped together based on the principal component to which they contribute (see \Sec{sec:pca}).}
\label{fig:obs_spearmans}
\end{figure*}

To test which of the different interpretations of ``dynamical state'' described in \Sec{sec:pca} are actually probed by X-ray and SZ morphological parameters, in \Fig{fig:obs_spearmans} we show the Spearman's rank correlation coefficient between each of these 12 observables (plus the composite parameters, $M_{\rm{X}}$ and $M_{\rm{SZ}}$) and the 17 dynamical state indicators that we use. Here we only show the magnitude of the Spearman's rank -- that is, we do not distinguish between positive and negative correlations, we are only studying the strength of the correlation.

Overall, the morphological parameters correlate well with dynamical state indicators that contribute strongly to the primary principal component, PC1, with a median correlation coefficient of 0.53 (median of the absolute values of the Spearman's rank coefficients). The exception to this is the Gaussianity of the cluster, which does not correlate as well with the PC1 parameters, regardless of whether it is measured from X-ray or SZ maps. This result is not unexpected -- previous studies have also found that Gaussianity is generally a less informative parameter than other X-ray/SZ morphological parameters. For example, \citet{deluca2021} show that $G$ is weighted three to four times weaker than the other parameters in the calculation of the composite parameter $M$. They attribute this to several factors: Gaussianity is mostly dependent on the global shape of a cluster, not on cluster substructure, and is also highly dependent on projection effects. Furthermore, even virialised dark matter haloes can have an ellipsoidal (non-spherical) shape, and so it is not necessarily a good metric for distinguishing relaxed and unrelaxed clusters. Similarly, \citet{cialone2018} show that Gaussianity calculated from SZ maps is less effective at separating clusters by dynamical state, and give similar reasons for this finding.

In both of these sets of mock observations, the weighted sums of morphological parameters ($M_{\rm{X}}$ and $M_{\rm{SZ}}$) correlate most strongly with the elements of PC1. This is predictable, given that $M$ is defined in such a way that the difference in $M$ between relaxed and unrelaxed clusters is maximised. These results are in general agreement with \citet{deluca2021}. It is important to note that the apertures for calculating these morphological parameters were chosen such that clusters could be separated based on $f_{\rm{s}}(R_{500})$ and $\Delta_{\rm{r}}(R_{500})$. However, the correlations of the morphological parameters with $f_{\rm{s}}(R_{200})$ and $\Delta_{\rm{r}}(R_{200})$ are similarly strong, and so we do not consider this to be an issue in our analysis.

The X-ray and SZ morphological parameters correlate far less well with the parameters that contribute to PC2, PC3 and PC4 -- the median (absolute) Spearman's rank coefficient is equal to 0.24 for these combinations. The exception to this is, once again, the cluster Gaussianity, particularly in X-rays. For parameters in PC1, the correlation coefficients associated with $G_{\rm{X}}$ were lower than those for the other morphological indicators, but for several parameters in PC2, PC3 and PC4, $G_{\rm{X}}$ is comparable to the other morphological indicators. For example, $G_{\rm{X}}$ has a moderate correlation with the cluster sphericity $c/a$ ($\rho_{\rm{s}}=0.23$, $p=2\times10^{-5}$), due to the fact that $G_{\rm{X}}$ is a direct measure of a cluster's shape.

These results are summarised in \Tab{tab:obs_vs_pca}, which shows how each of these 14 morphological measurements of each cluster correlate with the four main principal component values of that cluster. Indeed, we see that (apart from the Gaussianity), each of these X-ray/SZ measures correlates well with PC1, which we interpret as a measure of the time since much of the cluster's galaxy population was accreted and built up (see \Sec{sec:pca}). However, the halo virialisation (PC2), local environment (PC3) and recent merger history (PC4) do not correlate as well with any of the morphological parameters.

\begin{table}
	\centering
	\caption{Spearman's rank correlation coefficient between X-ray and SZ morphological parameters, and each of the four principal components. This data is similar to that shown in \Fig{fig:obs_spearmans}, but showing the correlation coefficient for each full principal component, not for their contributing dynamical state indicators.}
	\label{tab:obs_vs_pca}
	\begin{tabular}{ccccc} 
		\hline
		Observable & PC1 & PC2 & PC3 & PC4 \\
		\hline
        $A_{\rm{X}}$ &  -0.65 &  -0.11 &  -0.08 &   0.08\\
        $K_{\rm{X}}$ &   0.72 &   0.13 &  -0.07 &   0.04\\
        $W_{\rm{X}}$ &  -0.71 &  -0.17 &   0.04 &   0.06\\
        $P_{\rm{X}}$ &  -0.70 &  -0.03 &   0.01 &   0.02\\
        $G_{\rm{X}}$ &   0.25 &   0.11 &   0.02 &  -0.07\\
        $S_{\rm{X}}$ &  -0.57 &  -0.12 &  -0.04 &   0.06\\
        $M_{\rm{X}}$ &  -0.78 &  -0.13 &   0.01 &   0.05\\
        \hline
        $A_{\rm{SZ}}$ &  -0.64 &  -0.11 &  -0.04 &   0.11\\
        $K_{\rm{SZ}}$ &   0.75 &   0.23 &  -0.11 &   0.02\\
        $W_{\rm{SZ}}$ &  -0.73 &  -0.11 &  -0.01 &   0.07\\
        $P_{\rm{SZ}}$ &  -0.60 &  -0.01 &   0.02 &   0.03\\
        $G_{\rm{SZ}}$ &   0.18 &  -0.05 &  -0.05 &  -0.05\\
        $S_{\rm{SZ}}$ &  -0.63 &  -0.06 &  -0.08 &   0.12\\
        $M_{\rm{SZ}}$ &  -0.82 &  -0.14 &   0.02 &   0.06\\
		\hline
	\end{tabular}
\end{table}

\begin{figure*}
\includegraphics[width=\textwidth]{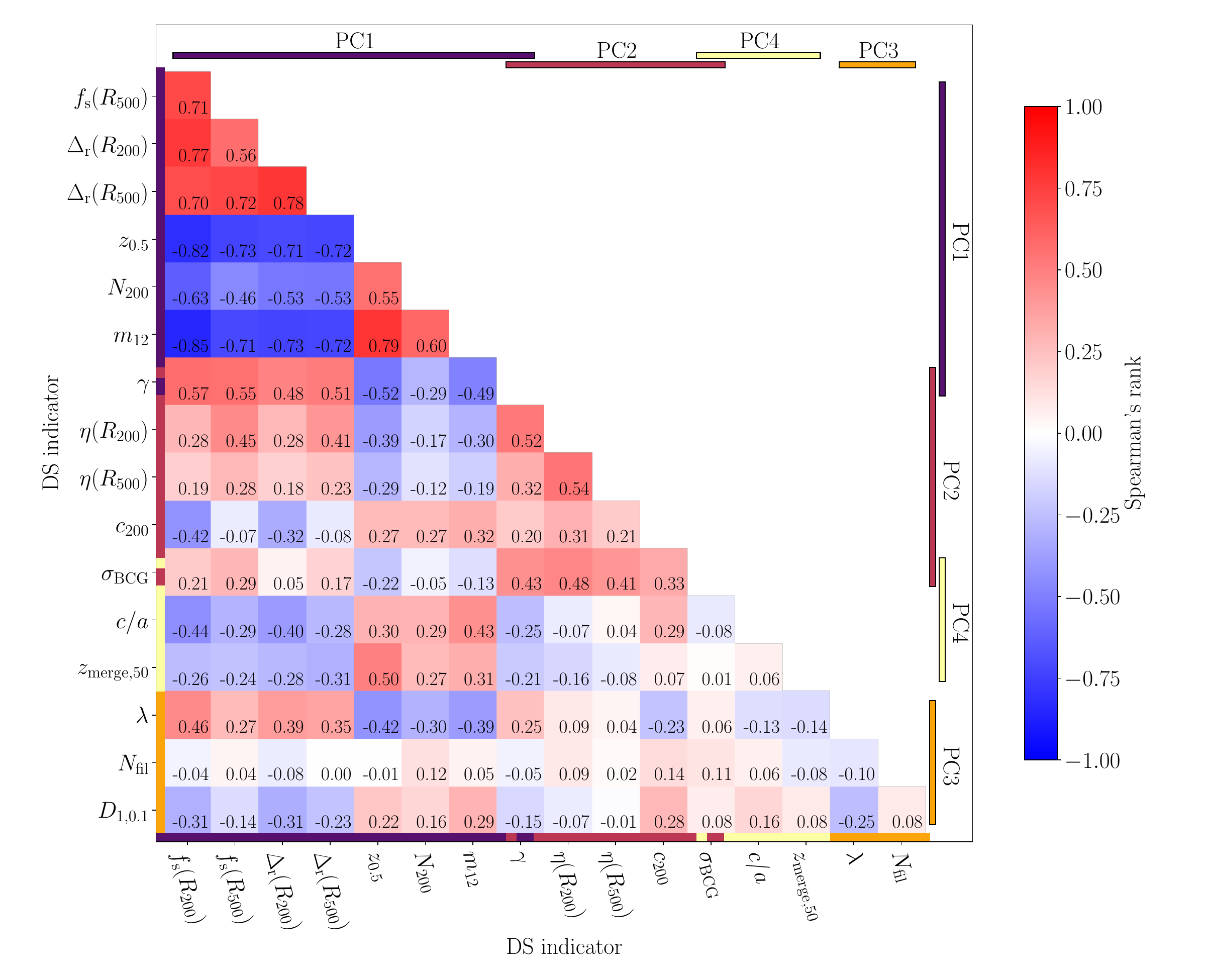}
\caption{Spearman's rank correlation coefficient, $\rho_{\rm{s}}$, between all 3D dynamical state parameters. Darker coloured cells represent stronger correlations; the value of the Spearman's rank is also written in each cell. A corner plot showing the distributions of each of these parameters can be found in \App{sec:ds_correlations}. The principal component to which each dynamical state indicator contributes is indicated at the top of the figure, and by the coloured bars along the axes.}
\label{fig:spearmans_3d}
\end{figure*}

Despite the fact that the observable morphological properties of a cluster only strongly correlate with the dynamical state indicators that make up PC1, there are still correlations between other indicators. \Fig{fig:spearmans_3d} shows the Spearman's rank correlation coefficient between all 17 of these 3D (non-observational) parameters. For example, the time since the last major merger, $z_{\rm{merge,50}}$, and the cluster formation time, $z_{\rm{0.5}}$, have a positive correlation ($\rho_{\rm{s}}=0.50$). This result is quite intuitive, as clusters that have recently merged will not have a high formation redshift. Other dynamical state indicators also correlate well -- for instance, all five of the parameters that contribute to PC2 show weak to moderate correlations with each other ($|\rho_{\rm{s}}|\geq0.20$, $p\leq2\times10^{-4}$). These correlations are weaker for the dynamical state indicators in PC3 and PC4, but this is not unexpected given that these components explain less of the variance of the total dataset (see \Fig{fig:explained_variance}).

This figure also displays some counter-intuitive results. For example, one might consider clusters with lots of substructure to be ``rich'', but there is actually a negative correlation between $f_{\rm{s}}$ and $N_{200}$ (for $f_{\rm{s}}(R_{200})$, $\rho_{\rm{s}}=-0.63$, and for $f_{\rm{s}}(R_{500})$, $\rho_{\rm{s}}=-0.46$). This is due to the fact that $N_{200}$ is defined by the number of galaxies of similar magnitude to the third brightest, and clusters with lots of substructure are more likely to have several bright galaxies, meaning that the threshold for galaxies to be counted in $N_{200}$ is higher. Nevertheless, $N_{200}$ is a useful dynamical state indicator, although using the term ``richness'' for this is somewhat ambiguous. While we do not explicitly show the correlations between these parameters in \Fig{fig:spearmans_3d} (for the sake of clarity), we have included a corner plot showing these correlations in \App{sec:ds_correlations}.

\subsection{Mass accretion histories of clusters}

In order to try and learn more about the dynamical histories of these clusters, we separate our sample of 324 clusters based on the coordinate values of their principal components, and study their mass accretion histories. A similar analysis was performed by \citet{wong2012}, who also split simulated dark matter haloes into different classes based on PCA analysis. They show that the mass accretion histories of their classes of clusters differ -- their first principal component separates the clusters into early-forming and late-forming, and their second component describes whether a halo's growth is accelerating or decelerating.

\begin{figure*}
\includegraphics[width=\textwidth]{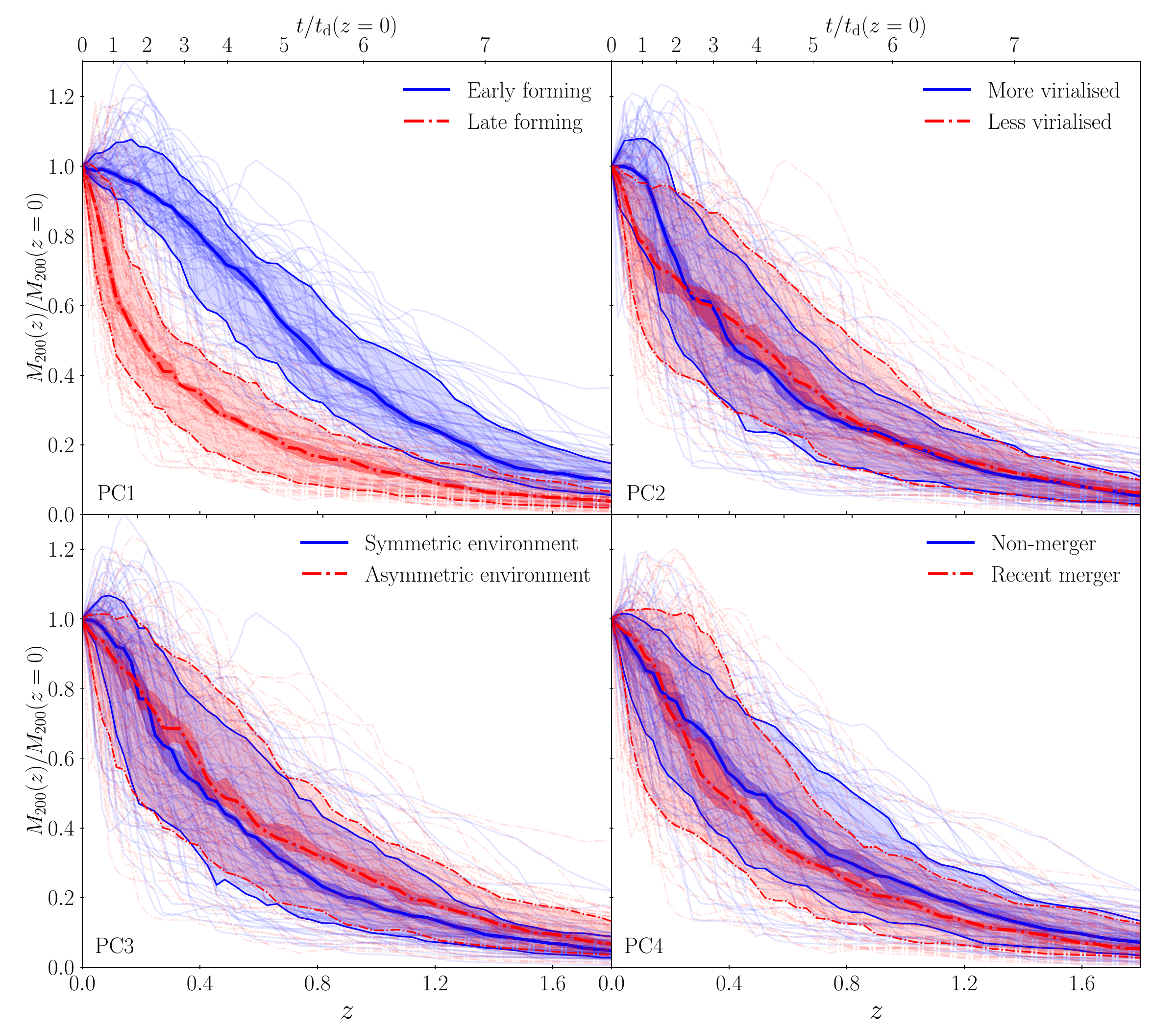}
\caption{Median mass accretion histories for clusters, split based on the values of their principal components, with a different principal component considered in each panel (PC1 in top-left, PC2 in top-right, PC3 in bottom-left, PC4 in bottom-right). Upper and lower quartile of each principal component are shown. Horizontal axis shows redshift (bottom axis), and lookback time in units of the dynamical time at $z=0$ (top axis), equal to approximately 1.4 Gyr as defined in \Eq{eq:dynamical_time}. Vertical axis shows the fraction of the $z=0$ mass that has been built up by a given time. Faint lines show the mass accretion histories of individual galaxy clusters. Some of these have temporary decreases in mass, which can be explained by several well-established mechanisms, such as the decrease in $M_{200}$ as a halo virialises after a period of rapid growth \citep[e.g.]{tormen1997}, or through mismatches in halo tracking \citep{fakhouri2009, behroozi2015}. The thick lines show the median accretion history for that group, the light shaded regions and thin lines show the $1\sigma$ (16-84 percentile) spread in the data, and the dark shaded regions show the uncertainty in the median.}
\label{fig:mah_pca_split}
\end{figure*}

For our similar analysis, we find the upper and lower quartiles of the values of each of our principal components. For example, based on PC1, we have a group of clusters that have low values of PC1, and a group with high values. As the contribution of $z_{0.5}$ to PC1 is positive (0.35, see \Tab{tab:prin_comps}), this also implies that selecting clusters with low values of PC1 is equivalent to choosing a group with low values of $z_{0.5}$, making them a ``late forming'' quartile. Similarly, we describe the group with high values of PC1 (and so higher values of $z_{0.5}$) as the ``early forming'' quartile. We then look at the median mass accretion history of the clusters in each of these two extreme groups, ignoring the intermediate clusters. The median mass accretion histories for each principal component are shown in \Fig{fig:mah_pca_split}.

When splitting clusters by their PC1 coordinate values, we see a clear difference between the mass accretion histories. As expected, the mass of the early forming clusters is built up far earlier. On average, the early forming clusters have built up half of their mass by a redshift of 0.7, while the late forming clusters do not do this until a redshift of 0.2. This is very similar to the first principal component found by \citet{wong2012}, which also splits their haloes into early-forming and late-forming.

We interpret the PC2 coordinate values as separating the clusters based on how virialised are their dark matter haloes. Separating the clusters based on their PC2 coordinate values, we find that the difference in their mass accretion histories is less pronounced than for PC1, and that there is a significant overlap in the spread of the data. However, by ``bootstrapping'' the data, we can study the uncertainty in the median mass accretion histories. From this, we find that there is a significant difference between the clusters split by PC2, particularly at low redshifts. The ``more virialised'' clusters have experienced only a very small change in mass since $z=0.1$, having an average mass at $z=0.1$ of $(0.97\pm0.01)M_{200}(z=0)$ (median and uncertainty). Meanwhile, the ``less virialised'' clusters have an average mass of $(0.80^{+0.02}_{-0.07})M_{200}(z=0)$ at $z=0.1$. This indicates that this ``virialisation'' component is strongly dependent on the very recent growth history of a cluster, although we emphasise that there is a large overlap in the spread of these data. We also note that the shapes of these mass accretion histories are similar to the mass accretion histories of haloes that \citet{wong2012} separate by PC2. They describe these as accelerating or decelerating growth rates, although their PC2 component is interpreted as describing the shape and spin of their haloes, not the ``virialisation'' as we find. Some slight variations between our results and those of \citet{wong2012} are to be expected, as their work uses dark matter-only simulations, not hydrodynamical simulations. The baryonic effects in \tth\ are particularly strong in the cluster centres \citep{haggar2021}, a region on which PC2 appears to be strongly dependent (see also \Sec{sec:conclusions}).

The component values of the third principal component, PC3, do not appear to have a strong impact on the mass accretion histories of clusters. Splitting the clusters by their coordinate values in this component (into ``symmetric environments'' with many filaments and a large distance to the nearest large halo, and ``asymmetric environments'' with few filaments and a nearby large halo) does not show a large difference in mass accretion histories. This indicates that the present-day local environment of a cluster is not closely tied to its mass accretion history, although \Fig{fig:spearmans_3d} does show a weak correlation between $z_{0.5}$ and $D_{1,0.1}$ ($\rho_{\rm{s}}=0.22$, $p=6\times10^{-5}$). We also note that a physical interpretation of this component is challenging. Based on past work, one would expect that relaxed clusters have both fewer filaments and a large distance to their nearest neighbour, but splitting clusters based on their PC3 values puts clusters with many filaments in a group with those that have a large distance to the nearest neighbour. However, this is likely due to the weak correlations between the components of PC3 that we find throughout this work, particularly $N_{\rm{fil}}$ (also shown in \Fig{fig:spearmans_3d}). We discuss the strength of these correlations further in \Sec{sec:umap}.

For the fourth principal component, PC4, the accretion histories are similar, although there is a slight difference in the shapes of the profiles; similarly to in PC1, the ``relaxed'' (non-merging) clusters build up their mass at slightly earlier redshifts. However, the main difference between the recent mergers and non-mergers is that the spread in mass accretion histories is far greater in recently merged clusters. For example, at $z=0.2$, the non-merging clusters have an average mass of $0.77^{+0.17}_{-0.23}$ times their present day mass (median and $1\sigma$ spread). In contrast, the recently merging clusters have an average mass of $0.73^{+0.28}_{-0.32}$ times their present day mass, corresponding to a $50\%$ greater spread. This increased spread in accretion histories for recently merging clusters is likely due to the stochastic nature of mergers -- these objects will have experienced a large jump in their mass at some recent time, but the exact time of this jump varies between clusters. Again, this is similar to the findings of \citet{wong2012}, who explain that mass accretion histories are not well modelled by smooth curves due to the stochasticity of merger events.

\subsection{Cluster dynamical state with UMAP}
\label{sec:umap}

Uniform manifold approximation and projection \citep[UMAP;][]{mcinnes2018} is an alternative dimensionality reduction technique to PCA. UMAP is commonly used as part of more complex machine learning studies, but is also a useful data visualisation method in its own right. UMAP involves first constructing a graph of datapoints (in our case, 324 galaxy clusters), each connected to their nearest $N$ neighbours in a high-dimensional space (the 17 dimensions defined by the dynamical state indicators). Some additional connections between points are also added, with a decreasing likelihood for points separated by greater distances. Next, a low-dimensional space, the ``embedding space'', is populated randomly with (324) corresponding datapoints. These are iteratively repositioned according to some loss function. This function is designed to match the pairwise distances between connected points in the new low-dimensional space to the pairwise distances in the original high-dimensional space. 

This process preserves local structure in the dataset, meaning that objects with similar properties remain close together in the low-dimensional ``embedding space''. It also preserves global properties of the dataset, so that groups of objects that are far apart in the high-dimensional space remain separated from one another. Similarly to PCA, UMAP is able to reduce a high-dimensional datasets into a smaller number of parameters (in our case, two). Unlike PCA however, the stochastic, iterative nature of UMAP allows for complex, non-linear relationships between parameters to also be captured. This means that the outputs of UMAP are harder to physically interpret -- whereas PCA produces mathematically well-defined axes, the output ``axes'' of UMAP are determined numerically. Similarly, a simple measure of the fractional variance that is captured by UMAP, analogous to the ``explained variance'' of PCA, does not exist \citep{mcinnes2018}. We use UMAP as a complementary method to PCA, approaching the same problem but in a different way.

\Fig{fig:umap} shows the results of our UMAP analysis for the 17 dynamical state indicators also used in the PCA. Each point represents one galaxy cluster, with the horizontal and vertical axes showing the two combined parameters produced by the two dimensional UMAP analysis, analogous to the principal components that we summarised in \Tab{tab:prin_comps}. Due to the definition and optimisation processes in UMAP, points that are close together on a UMAP plot are close together in the 17-dimensional space that we started with. Consequently, these are likely to be dynamically similar clusters. As the UMAP axes are a dimensionless (due to the standardisation of the data), complex combination of many parameters, these axes are left unlabelled for clarity.

Each panel in this figure shows data for one of our 17 3D dynamical state indicators, and the colour of each point in this panel represents the value of this dynamical state indicator for each cluster. Linear colour scales are used for each, but the colour scale is flipped for some quantities, such that the clusters we interpret as more dynamically relaxed (according to this individual parameter) are shown by darker colours. This allows us to see which regions of this new UMAP embedding space contain dynamically relaxed clusters, according to each of the different definitions. Similarly to in \Fig{fig:obs_spearmans}, the panels are grouped by the principal component to which they most strongly contribute, according to our principal component analysis -- otherwise, this approach is fully independent of our PCA.

For each of these panels, we calculate the direction in which the $z$-axis (i.e. the value of the dynamical state indicator) varies most quickly, and thus the direction in this space in which the clusters become more dynamically unrelaxed. We do so using the partial correlation coefficients (PCCs) approach described in \citet{lawrance1976}; see also \citet{baker2022} for a clear explanation of this method. The partial correlation coefficient between two quantities, $A$ and $B$, whilst controlling for a third quantity, $C$, is given by $\rho_{AB|C}$, and is calculated using the equation below:
\begin{equation}
    \rho_{AB|C}=\frac{\rho_{AB}-\rho_{AC}\rho_{BC}}{\sqrt{1-\rho_{AC}^{2}}\sqrt{1-\rho_{BC}^{2}}}\,.
    \label{eq:pcc}
\end{equation}

Here, $\rho_{XY}$ is the Spearman's rank correlation coefficient between two quantities, $X$ and $Y$. \Eq{eq:pcc} allows the partial correlation coefficent between the colour (the value of the dynamical state indicator) and the vertical UMAP axis to be found, whilst controlling for the horizontal axis. Similarly, we find the PCC between the colour and the horizontal UMAP axis, whilst controlling for the vertical axis. The ratio of these two partial correlation coefficients can then be used to calculate the maximal variation direction of the colour in the UMAP embedding space \citep[see also][for further details of this method]{bluck2020}. We plot each of these directions of maximal dynamical state variation in the top-left panel of \Fig{fig:umap}, coloured by the principal component to which they contribute most strongly. The length of each arrow is equal to the Spearman's rank correlation coefficient between the dynamical state indicator (colour) and the position of clusters along this direction in embedding space. Each arrow is also shown on the bottom-left corner of the panel to which it corresponds. This top-left panel therefore shows which direction in this embedding space contains the unrelaxed clusters, according to each of the 3D dynamical state measures. 

\afterpage{
\begin{landscape}
\begin{figure}
\vspace{28pt}
\includegraphics[width=\columnwidth]{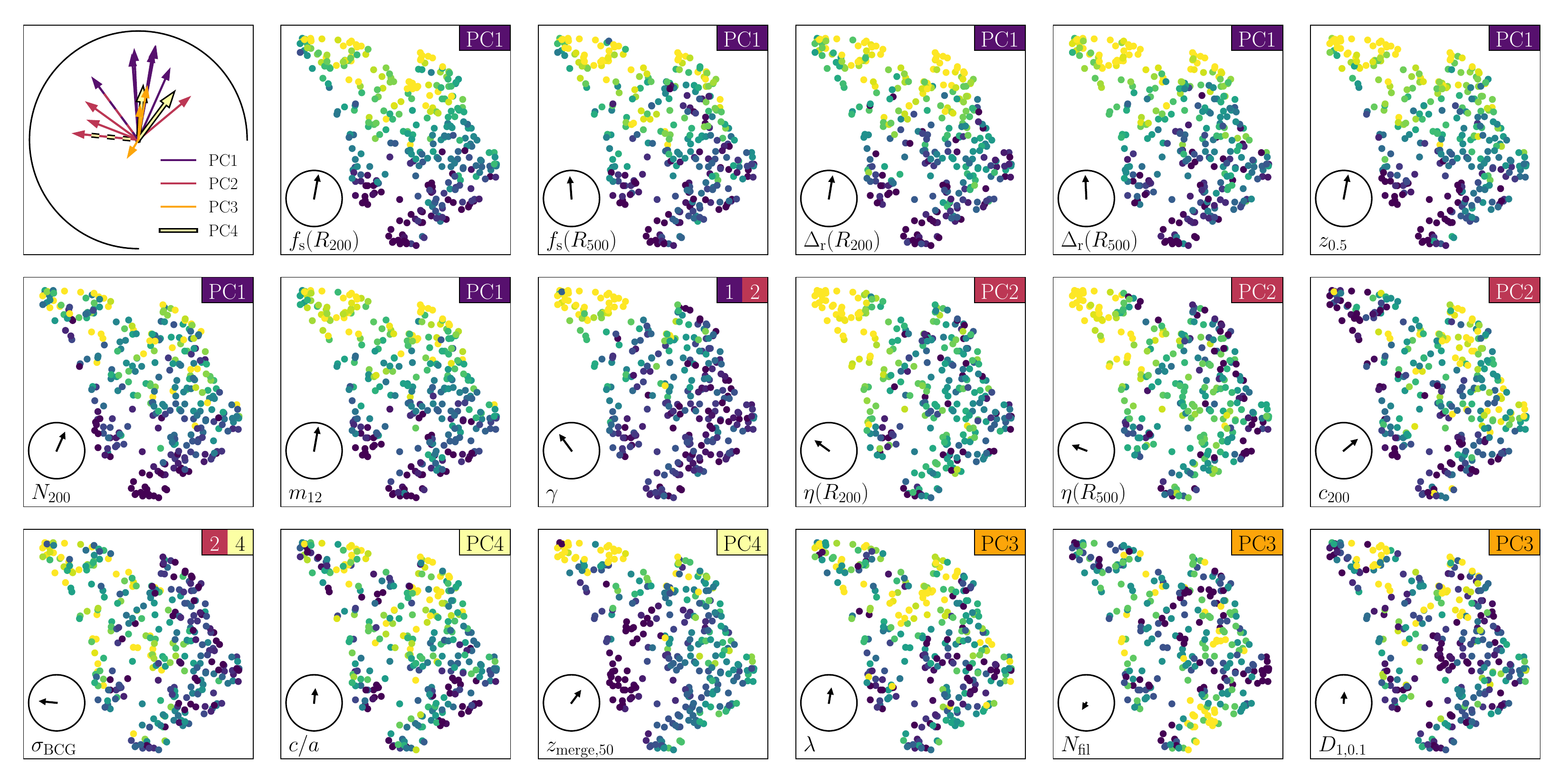}
\caption{Uniform manifold approximation and projection (UMAP) of 17 cluster properties, also used in principal component analysis (PCA). For all panels apart from top-left, each point represents one galaxy cluster, with the horizontal and vertical axes representing the first and second UMAP axes, respectively. The colour of each point shows the value of one dynamical state indicator for that cluster -- the specific indicator is written in the bottom-left of each panel. The colour scale has been implemented using our interpretation of the dynamical state parameters, such that a darker colour represents clusters that are dynamically relaxed according to that single parameter. For $f_{\rm{s}}(R_{200})$, $f_{\rm{s}}(R_{500})$, $\Delta_{\rm{r}}(R_{200})$, $\Delta_{\rm{r}}(R_{500})$, $\gamma$, $\eta(R_{200})$, $\eta(R_{500})$, $\sigma_{\rm{BCG}}$, $\lambda$ and $N_{\rm{fil}}$, lighter colours represent greater values of these parameters (as defined in \Sec{sec:3d_measures}). For $z_{0.5}$, $N_{200}$, $m_{12}$, $c_{200}$, $c/a$, $z_{\rm{merge,50}}$ and $D_{1,0.1}$, lighter colours represent lower values. For simplicity, quantitative colour bars are not shown here. In the top-right of each panel, the principal component (or components, in the case of $\gamma$ and $\sigma_{\rm{BCG}}$) to which that parameter belongs is indicated. The dynamical state indicators are grouped in the same way as in \Fig{fig:obs_spearmans}. For each of these panels, we calculate the direction in which the clusters become more disturbed (less relaxed) using partial correlation coefficients (PCC) analysis. These directions, coloured by the principal component to which the parameter belongs, are shown in the top-left panel. For $\gamma$ and $\sigma_{\rm{BCG}}$, these arrows are dashed with colours corresponding to the two components to which they contribute. The length of each arrow represents the strength of the correlation, quantified by the Spearman's rank correlation coefficient, $\rho_{\rm{s}}$, with the black circle showing where $\rho_{\rm{s}}=1$. The magnitude and directions of each arrow is also included in the bottom-left of the panel to which it corresponds. With the exception of $N_{\rm{fil}}$, all of these correlations are significant ($p\leq10^{-5}$).}
\label{fig:umap}
\end{figure}
\end{landscape}}

\Fig{fig:umap} shows that each of the parameters making up the first principal component of dynamical state ($f_{\rm{s}}(R_{200})$, $f_{\rm{s}}(R_{500})$, $\Delta_{\rm{r}}(R_{200})$, $\Delta_{\rm{r}}(R_{500})$, $z_{0.5}$, $m_{12}$ and $N_{200}$) vary in approximately the same direction across the UMAP plots, from bottom to top. According to these measures, the relaxed clusters are generally found at the bottom of these plots, and the unrelaxed clusters at the top. This indicates that the UMAP has also grouped clusters based on this interpretation of dynamical state, separating clusters that have formed and developed their substructure recently, from those that formed long ago. The top-left panel of \Fig{fig:umap} confirms that the increasingly disturbed clusters, according to PC1 parameters, are at the top of the UMAP plots.

However, the dynamical state indicators that contribute to the second, third and fourth principal components do not all vary in the same way. The PC4 parameters ($c/a$ and $z_{\rm{merge,50}}$) do follow a similar trend to those in PC1, with more dynamically disturbed clusters (low-$z_{\rm{merge,50}}$ and low-$c/a$) being found near the top of the plot -- this is consistent with our interpretation of PC4 in \Sec{sec:pca}, where we described how some of the merger history behaviour is also included in PC1. The PC2 indicators, representing the virialisation and concentration of the cluster haloes ($\eta(R_{200})$, $\eta(R_{500})$ and $c_{200}$) instead vary more strongly from right to left, although not all in the same direction; $\gamma$ follows a similar trend. The UMAP can also separate clusters by their environmental dynamical state indicators (PC3: $\lambda$, $N_{\rm{fil}}$ and $D_{1,0.1}$), but not very well, indicated by the weak trends in these panels. The high-$\lambda$ and low-$D_{1,0.1}$ clusters, usually interpreted as unrelaxed, are located towards the top of the plot -- these trends are not strong, but are significant, and are along the same axis as the PC1 components' trends. For spin, $\lambda$, $|\rho_{\rm{s}}|=0.44$. For the environment parameter, $D_{1,0.1}$, $|\rho_{\rm{s}}|=0.24$ ($p=10^{-5}$). The number of filaments and UMAP position correlate in the opposite direction along the same axis, but this correlation is not significant ($|\rho_{\rm{s}}|=0.10$, $p=0.06$).

This UMAP analysis gives many subtle results that are not immediately straightforward to interpret, but overall it confirms the indications of our PCA -- the different groupings of clusters in the plots correspond to different classes of dynamically relaxed objects. For example, the early-forming, low-substructure clusters (PC1) are found at the bottom of the UMAP plots. These clusters have highly-concentrated haloes, as would typically be expected of a relaxed cluster \citep{yuan2020}. At the top of the UMAP plots, we instead find recently-formed clusters. However, there is also a third major population of clusters in the top-left of the plots -- these are dynamically unrelaxed haloes, that have formed recently, have much substructure and have non-virialised haloes, yet are highly-concentrated. This population have recently experienced major mergers ($z_{\rm{merge,50}}$), and have very high accretion rates, $\gamma$, indicating that they are currently accreting large amounts of material. This is in contrast to the other clusters at the top of these plots, which have recently merged but have not accreted lots of material in the last dynamical time (since \mbox{$z=0.1$}). This difference is equivalent to the mass accretion histories in the top-right panel of \Fig{fig:mah_pca_split}, where we group clusters based on PC2 and show that their recent ($z<0.2$) growth histories are different.

These three populations can be interpreted as follows: high-concentration galaxy clusters are likely to have formed long ago, while low-concentration clusters are likely recently formed, often after a major merger. However, if a merger is still ongoing, or if large amounts of galaxies are pulled into the cluster immediately afterwards (potentially accompanying the merging cluster), this can make a cluster appear highly-concentrated, and therefore dynamically relaxed. This explanation is strongly supported by \citet{wang2020}, who show that major mergers can produce oscillations in concentration, driving concentration up significantly when merging material reaches the first pericentre of its orbit, before concentration quickly decreases again. Alternatively, if a cluster is continuing to accrete large amounts of diffuse material into its halo (rather than a single large object), this could also result in a high mass accretion rate without a corresponding decrease in halo concentration. 

As a result of this, a simple measurement of the concentration of a cluster is insufficient to draw conclusions about the virialisation of its halo, its formation time, or its merger history. Additionally, the dynamical state of a cluster's BCG ($\sigma_{\rm{BCG}}$) also seems to only probe some of these unrelaxed clusters. This complex behaviour cannot be fully accounted for by a 1D linear fit, which explains our previous counter-intuitive result that highly-concentrated clusters appear to be less virialised and rapidly-accreting (\Fig{fig:pca_bars} and \Fig{fig:spearmans_3d}); simplifying this to a single correlation coefficient does not capture the complete behaviour.

Additionally, the UMAP analysis shows that the early-forming clusters can be split into two groups based on their merger histories. The group in the bottom-left have not experienced major mergers for a very long time ($\sim10$ Gyr), or not at all throughout their history. Meanwhile, those in the bottom-right have experienced a major merger more recently, although still not for several gigayears. There seems to be a (weak) correlation with the connectivity of these clusters, $N_{\rm{fil}}$: early-forming clusters that have still experienced a major merger are more strongly connected than early-forming clusters that have never experienced a major merger. This may be indicative of the extremely long timescales ($\sim10$ Gyr) over which cosmic filaments are persistent, but a more detailed analysis would be needed to investigate this connection.

\section{Discussion and Conclusions}
\label{sec:conclusions}

Using hydrodynamical simulations of galaxy clusters, we show that the wide variety of parameters used in the literature to quantify the ``dynamical state'' of a cluster are actually probing multiple properties of a cluster. Consequently, by applying PCA, we conclude that the dynamical state of a cluster as described in the literature is actually made up of approximately four different properties, summarised in \Sec{sec:pca}. To use any measurement (or set of measurements) to describe a cluster as simply ``dynamically relaxed'' or ``dynamically unrelaxed'' is an incomplete description. Different classes of dynamically relaxed clusters exist, and so instead one must specify in which sense is a cluster relaxed. 

The main component of a cluster's dynamical state is its ``formation dynamical state'', describing the formation time of a cluster -- that is, whether a cluster built up most of its mass and galaxy population recently or long ago. This formation state is indicated by galaxies and substructure within the cluster. Recently forming clusters have more substructure and a greater offset between their centre-of-mass and the position of their BCG. Additionally these clusters are typically less dominated by a single, bright galaxy, and are accreting material at a faster rate. This ``formation state'' is similar to the dynamical state used in many previous studies \citep[for example]{cui2017, gouin2021}. It should be noted that several such studies include the virial ratio, $\eta$, in this measure of dynamical state, making it analogous to a combination of our PC1 and PC2, however it has been shown that the virial ratio is not an important contributor to this measure of dynamical state \citep{haggar2020,deluca2021}. As we see in \Fig{fig:mah_pca_split}, separating clusters based on their first principal component from our PCA is equivalent to separating them based on their formation times, in agreement with \citet{wong2012}. The ``late-forming'' clusters have a median formation time of \mbox{$z_{0.5}=0.2\pm0.1$}, compared to \mbox{$z_{0.5}=0.7\pm0.2$} for the relaxed ``early-forming'' clusters.

In addition to this dynamical description of the subhaloes and galaxies in a cluster, the diffuse material in the cluster halo itself can be virialised and dynamically relaxed. This ``virialisation dynamical state'' describes how well the material in a cluster obeys the modified virial theorem \citep[accounting for surface pressure, see][]{poole2006, shaw2006}. It is particularly well-described by the properties of the central regions of a cluster -- the NFW concentration of the halo, and the velocity dispersion of the brightest cluster galaxy -- as well as the present day accretion rate of the cluster, $\gamma$. The UMAP analysis in \Sec{sec:umap} also showed that, in this space, the ``virialisation state'' varies along a different axis to the ``formation state'', described by PC1. These complex relationships can be seen through the apparent inconsistencies when only considering monotonic relationships between quantities. For example, \Fig{fig:spearmans_3d} shows that highly-concentrated haloes (typically considered ``relaxed'') correlate with having less substructure (a sign of being relaxed), but also with a less virialised halo according to $\eta$. The complex relationships between these dynamical state indicators are much more apparent in \Fig{fig:umap}.

Clusters can also be dynamically relaxed in terms of their local environment, or in terms of their merger history. These components of dynamical state are driven by global properties of a cluster, such as their shape and spin. These components are typically noisier, and the cluster properties that constitute these components do not correlate as well with one another. However, it is still meaningful to separate clusters based on this -- for instance, \Fig{fig:mah_pca_split} shows an increased spread in the mass accretion histories of recently-merged clusters.

Of these four forms of dynamical state, X-ray and SZ measurements of morphological properties of clusters, such as their asymmetry, concentration, and centroid shift, overwhelmingly probe the formation dynamical state (PC1). Generally, X-ray and SZ morphological parameters are very well-suited to describing the formation of clusters and their accretion of substructures, but not well-suited to describing the virialisation of cluster haloes, their merger history, or their cosmic environment. This is not unexpected: as we discussed in \Sec{sec:umap}, mergers appear to have a somewhat chaotic effect on other dynamical state parameters, and can lead to a cluster halo appearing either relaxed or unrelaxed. Additionally, these morphological measurements, particularly the X-ray measurements, are heavily weighted towards the central regions of clusters ($R_{500}$ or smaller). We would therefore not necessarily expect them to be good indicators of clusters' larger-scale cosmic environments. There are also some slight exceptions to this rule -- for example, the light concentration ratio calculated from the SZ-effect, $K_{\rm{SZ}}$, does indeed correlate with the NFW concentration, $c_{200}$ ($\rho_{\rm{s}}-0.37$).

As stated previously, this work is very much a theoretical, simulation-focused study of dynamical state, and any further detailed analysis of observable cluster properties is beyond the scope of this paper. However, numerous other observable properties of clusters exist, and in future work we hope to examine how these observable quantities correspond to dynamical state. For example, although the mock X-ray observations used in this work are focused on the cluster centres, X-rays can also be used to map out substructures in cluster outskirts \citep{zhang2020}, and even cosmic filaments \citep{walker2019, biffi2022}. Other measures of cluster X-ray and SZ maps also exist, such as decomposition into Zernike polynomials \citep{capalbo2021}. Beyond this, optical data can also be used to quantify the dynamical states of clusters, by determining properties such as their substructure \citep{wen2013}, richness, and brightest galaxies magnitude difference, $m_{12}$ \citep{zhoolidehhaghighi2020}. A future study (Cornwell et al., in prep.) will use \tth\ simulations to investigate how cluster dynamical states can be determined using spectroscopic measurements.

The results from this paper confirm that, although a description of dynamical state \citep[see][]{binney1987} is theoretically quite simple, the dynamical state of galaxy clusters in practice is more complex. This has implications for wider work on galaxy clusters. Many studies split clusters into two samples, ``relaxed'' and ``unrelaxed'', based on a small number of properties, but this has the potential to combine clusters with very different dynamical histories into a single group. Instead, it is important to describe in which sense a cluster is known to be dynamically relaxed or unrelaxed -- for example, in terms of their substructure accretion history, their recent merger history, or how virialised they are at the present day.

In future we plan to carry out a more quantifiable analysis of this multi-dimensional dynamical state, rather than the qualitative description presented in this work. Instead of a linear scale of dynamical state, it may be more natural to quantify the dynamical state of clusters in two or more dimensions; the UMAP analysis in this paper is similar to this, but it not simple to interpret physically. Such a study could also be extended to study the dynamical state in different ``apertures'', looking at the kinematics of the cluster centre, or of its outskirts. Additionally, we plan to investigate whether clusters can be naturally separated into groups based on their dynamical states. Previous work \citep[e.g.][]{zhang2022} has investigated defining a bimodal function to describe dynamical state, to allow clusters to be split into ``relaxed'' and ``unrelaxed'' groups. In a future study, we will investigate whether a multimodal description of dynamical state exists, that would allow clusters to be split into more than two groups. This would provide a means to compare the properties of clusters that do actually have similar dynamical states, and to select samples of truly relaxed galaxy clusters in order to reduces biases in cosmological studies.

\section*{Acknowledgements}

We would like to thank the anonymous referee for their comments on this work, which have helped to improve the paper by adding robustness and clarity.

This work has been made possible by \threehun\ collaboration\footnote{\url{https://www.the300-project.org}}. This work has received financial support from the European Union's Horizon 2020 Research and Innovation programme under the Marie Sk\l{}odowskaw-Curie grant agreement number 734374, i.e. the LACEGAL project\footnote{\url{https://cordis.europa.eu/project/rcn/207630\_en.html}}. \threehun\ simulations used in this paper have been performed in the MareNostrum Supercomputer at the Barcelona Supercomputing Center, thanks to CPU time granted by the Red Espa\~nola de Supercomputaci\'on. 

WC is supported by the STFC AGP Grant ST/V000594/1 and the Atracci\'{o}n de Talento Contract no. 2020-T1/TIC-19882 granted by the Comunidad de Madrid in Spain. He also thanks the Ministerio de Ciencia e Innovaci\'{o}n (Spain) for financial support under Project grant PID2021-122603NB-C21 and ERC: HORIZON-TMA-MSCA-SE for supporting the LACEGAL-III project with grant number 101086388. M.D.P. acknowledges financial support from PRIN 2022 (Mass and selection biases of galaxy clusters: a multi-probe approach - n. 20228B938N) and from Sapienza Universit\`{a} di Roma - Progetti di Ricerca Medi 2022, RM1221816758ED4E. JET acknowledges support from an NSERC Discovery Grant.

This work makes use of the {\sc{SciPy}} \citep{virtanen2020}, {\sc{NumPy}} \citep{vanderwalt2011}, {\sc{Matplotlib}} \citep{hunter2007}, {\sc{pandas}} \citep{mckinney2010} and {\sc{Scikit-learn}} \citep{pedregosa2011} packages for {\sc{Python}}\footnote{\url{https://www.python.org}}.

The authors contributed to the paper in the following ways: RH carried out the analysis, wrote the paper, and produced the figures. FDL and MDP calculated the morphological parameters from the X-ray/SZ mock observations, which were produced by WC. ES carried out the UMAP analysis. AK produced the simulation halo catalogues and merger trees, from which several of the 3D dynamical state measures were calculated. AK also assisted with the general direction of the work, along with JET, MEG and FRP. ACS calculated the times since last major merger. WC calculated the dynamical state indicators $f_{\rm{s}}$, $\Delta_{\rm{r}}$ and $\eta$. UK calculated the cluster connectivities. 
RAMP calculated the cluster formation times, $z_{0.5}$. CP calculated the BCG velocity dispersions. All authors had the opportunity to comment on the paper.

\section*{Data Availability}

The simulations underlying this work have been provided by \threehun\ collaboration. The data may be shared on reasonable request to the corresponding author, with the permission of the collaboration.



\bibliographystyle{mnras}
\bibliography{main} 



\appendix

\section{Explained variance of principal components}
\label{sec:explained_variance}

\begin{figure}
\includegraphics[width=\columnwidth]{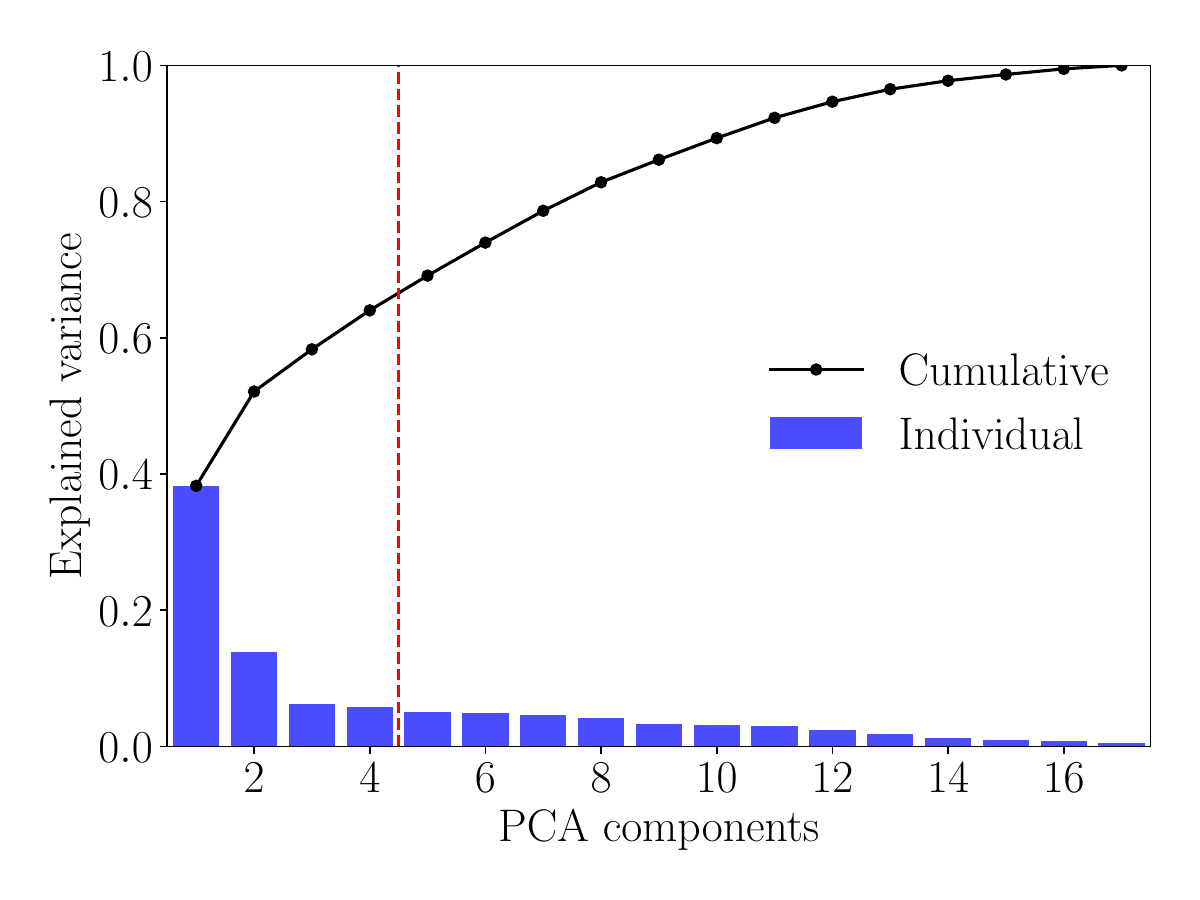}
\caption{Cumulative proportion of variance of the dynamical state indicators dataset explained by the ordered principal components (black line and points), and by the individual components (blue bars). Throughout our analysis, we only consider the first four components, indicated by the vertical red dashed line.}
\label{fig:explained_variance}
\end{figure}

The four principal components selected and analysed throughout this work collectively describe $64\%$ of the total variance of the 17 dynamical state indicators described in \Sec{sec:3d_measures}. In \Fig{fig:explained_variance}, we show the proportion of the variance explained by all 17 of the principal components, and the cumulative explained variance. 

Clearly, PC1 is the dominant component, explaining $38\%$ of the total variance alone. Other than this, there is not a clearly visible ``cut-off'' point, beyond which parameters are far less important. Consequently, the decision to keep only the first four principal components is not obviously mathematically justifiable. This choice was made in order to select the minimum number of components such that each of the dynamical state indicators contributed strongly to at least one of them, to provide an idea of the number of dimensions along which dynamical state varies. For example, PC5 (which explains $5\%$ of the total variance) is dominated by $\lambda$, $N_{\rm{fil}}$ and $D_{1,0.1}$, which all also strongly contribute to PC3 (see \Tab{tab:prin_comps_pc5}). We interpret this as PC5 providing a ``second order'' correction to the environment of a cluster, as described by PC3; we believe that including such a component would not meaningfully add to this work.

\begin{table}
	\centering
	\caption{Similar to \Tab{tab:prin_comps}, but showing coordinate values for PC5, which is excluded throughout this work. Component coordinates with an absolute value greater than 0.24 are highlighted in bold; the three parameters that contribute strongly to PC5 ($\lambda$, $N_{\rm{fil}}$ and $D_{1,0.1}$) also contribute strongly to PC3, albeit with different weights.}
	\label{tab:prin_comps_pc5}
	\begin{tabular}{cc} 
		\hline
		Parameter & Contribution to PC5 \\
		\hline
        $f_{\rm{s}}(R_{\rm{200}})$ &  -0.04\\
        $f_{\rm{s}}(R_{\rm{500}})$ &   0.05\\
        $\Delta_{\rm{r}}(R_{\rm{200}})$ &  -0.02\\
        $\Delta_{\rm{r}}(R_{\rm{500}})$ &   0.02\\
        $\eta(R_{\rm{200}})$ &  -0.14\\
        $\eta(R_{\rm{500}})$ &  -0.13\\
        $z_{\rm{0.5}}$ &  -0.04\\
        $\lambda$ & \bf{ -0.40}\\
        $c/a$ &  -0.23\\
        $c_{\rm{200}}$ &   0.08\\
        $z_{\rm{merge,50}}$ &  -0.14\\
        $\gamma$ &   0.09\\
        $N_{\rm{fil}}$ & \bf{ -0.55}\\
        $D_{1,0.1}$ & \bf{  0.59}\\
        $N_{200}$ &  -0.24\\
        $m_{12}$ &  -0.05\\
        $\sigma_{\rm{BCG}}$ &   0.05\\
        \hline
	\end{tabular}
\end{table}

\section{Correlations between dynamical state indicators}
\label{sec:ds_correlations}

In \Fig{fig:spearmans_3d} we show the Spearman's rank correlation coefficient between each of the 3D dynamical state indicators used in this study (described in \Sec{sec:3d_measures}). In \Fig{fig:ds_corner_plot} below, we explicitly show the correlations between these measures, from which these correlation coefficients were calculated. As in \Fig{fig:spearmans_3d}, we also group these parameters based on the principal component to which they most strongly contribute. Finally, we also include the cluster mass as an additional parameter here, to explicitly show the mass dependence of these quantities.

\begin{figure*}
\includegraphics[width=\textwidth]{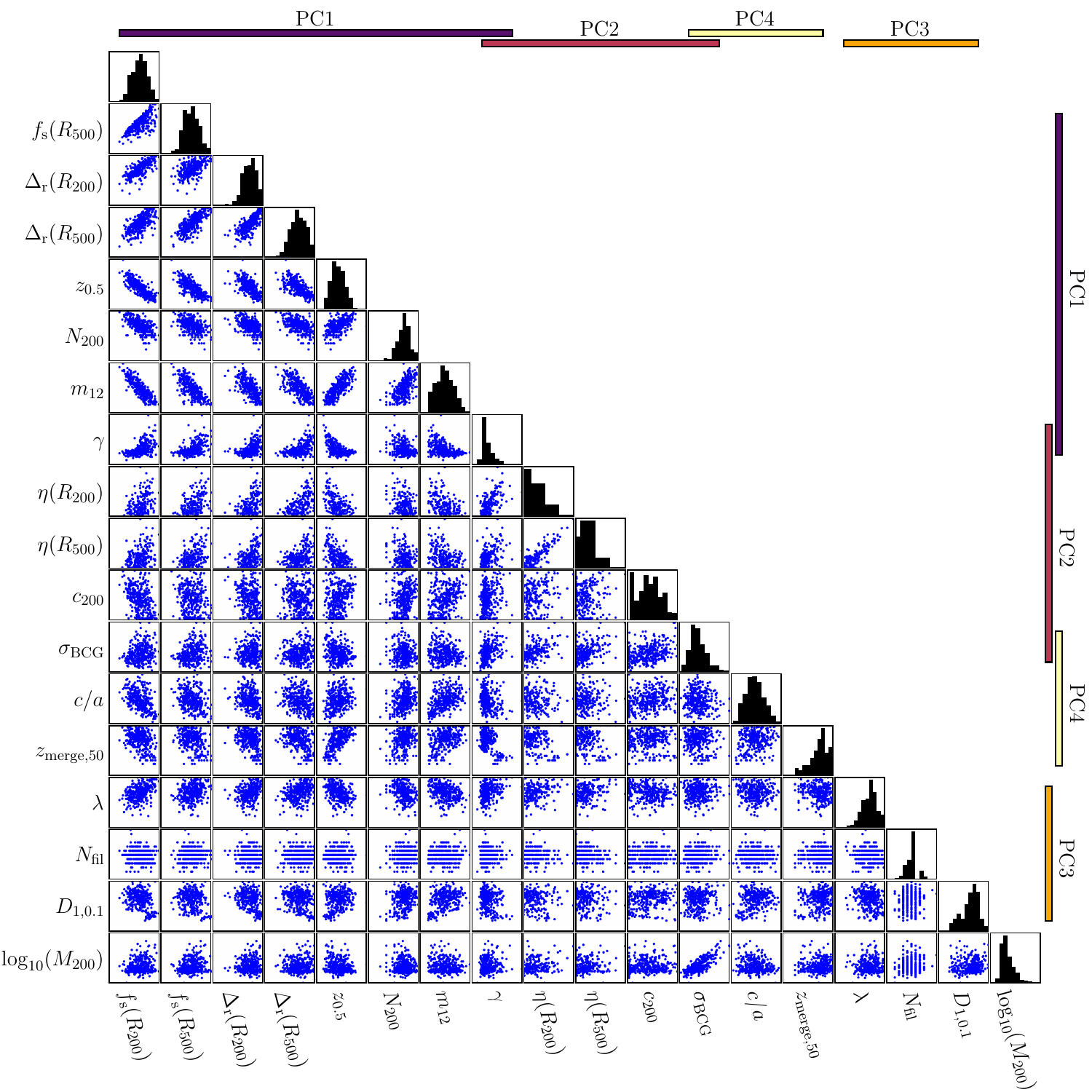}
\caption{Corner plot showing the correlations between the 17 dynamical state indicators described in \Sec{sec:3d_measures}. The Spearman's rank correlation coefficients displayed in \Fig{fig:spearmans_3d} are calculated from these scatter plots. This figure also includes histograms showing the distribution of these parameters. We also include the logarithmic mass of each galaxy cluster as an additional quantity here, to show any mass dependence of these 17 parameters.}
\label{fig:ds_corner_plot}
\end{figure*}

It should be noted that several of these parameters have a highly skewed distribution (for example, $\eta$). These parameters were all standardised for the PCA and UMAP analysis in this work to approximate a normal distribution with a mean of zero and a unit standard deviation, but here we show the raw distributions. Most of these parameters also display either no dependence, or a weak dependence, on cluster mass ($|\rho_{\rm{s}}|\leq0.15$, $p\geq0.005$). The exception is the velocity dispersion of the BCG ($\rho_{\rm{s}}=0.74$), which has a fairly strong positive correlation with the cluster mass. This result is expected, as numerous previous studies have found that the BCG velocity dispersion scales with the mass of its host cluster \citep[e.g][]{sohn2020}. One could account for this mass dependence by normalising the BCG velocity dispersion by the maximum circular orbital speed around the cluster, $v_{\rm{circ}}$. For \tth, the Spearman's rank correlation coefficient between $M_{200}$ and $\sigma_{\rm{BCG}}/v_{\rm{circ}}$ is far reduced ($\rho_{\rm{s}}=0.19$, $p=6\times10^{-4}$); although we have chosen to just use $\sigma_{\rm{BCG}}$ in this study, this ratio is a potential alternative choice that one could use instead.


\bsp	
\label{lastpage}
\end{document}